# Carbon Nanotube-Based Black Coatings

J. Lehman, C. Yung, N. Tomlin, D. Conklin, and M. Stephens

*Applied Physics Division, National Institute of Standards and Technology, Boulder, Colorado, 80305, USA*

Coatings comprised of carbon nanotubes are very black; that is, characterized by low reflectance over a broad wavelength range from the visible to far infrared. Arguably there is no other material that is comparable. This is attributable to the intrinsic properties of graphene as well as the morphology (density, thickness, disorder, tube size) of the coating. The need for black coatings is persistent for a variety of applications such as baffles and traps for space instruments. Because of the thermal properties, nanotube coatings are also well suited for thermal detectors, blackbodies and other applications where light is trapped and converted to heat. We briefly describe a history of other coatings such as nickel phosphorous, gold black and carbon-based paints and the comparable structural morphology that we associate with very black coatings. In many cases, it is a significant challenge to put the blackest coating on something useful. We describe the growth of carbon nanotube forests on substrates such as metals and silicon along with the catalyst requirements and temperature limitations. We also describe coatings derived from carbon nanotubes and applied like paint. Another significant challenge is that of building the measurement apparatus and determining the optical properties of something having negligible reflectance. There exists information in the literature for effective media approximations to model the dielectric function of vertically aligned arrays. We summarize this as well as other approaches that are useful for predicting the coating behavior along with the refractive index of graphite from the literature that is necessary for the models we know of. In our experience, the scientific questions can be overshadowed by practical matters, so we provide an appendix of our best recipes for making as-grown, sprayed or other coatings for the blackest and most robust coating for a chosen substrate and a description of reflectance measurements.

## I. INTRODUCTION:

The nature of the blackest coatings relies not only on the intrinsic properties of the material, but also the morphology and light-trapping ability. The physics of such coatings can be described with classical physics and intuitively by considering a medium having an index of near unity such that incoming light is not immediately reflected and with sufficient depth of structure, photons are eventually absorbed and converted to heat. Metal blacks, or metal soot derived from pure gold or pure silver illustrate this point. Such coatings have been known for nearly one hundred years.[1] One can begin with something that is intrinsically highly reflective, like gold, and make it extremely black by piling up atoms into a structure that appears as cotton wool at the micrometer scale. The earliest thermal detectors for studying the climate relied on candle soot to absorb incoming radiation.[2] The need for black coatings applies to efficiently absorbing optical and infrared radiation to capture the energy primarily and secondly to prevent the light from going elsewhere. Carbon soot is still considered a low-cost and viable black absorber for large-area surfaces and it is apparent that the morphology may be correlated with the blackness of the coating, which may be correlated with the raw material from which the soot is derived.[3]

During the past ten years, black coatings derived from carbon nanotubes have been investigated. Hata et al., demonstrated large uniform arrays of carbon nanotubes by the method of water assisted growth in 2004.[4] The first investigation by Lehman et al., was a sprayed coating of highly purified single-wall carbon nanotubes on a pyroelectric detector.[5] This coating was characterized by the apparent manifestation of what we now know of as excitonic intraband transitions of single-wall carbon nanotubes.[6] In 2006, this was followed by multiple attempts to grow vertically aligned carbon nanotubes on a pyroelectric detector substrate.[7,8] This coating showed great promise for being spectrally uniform over a broad wavelength range. In 2007, Yang and co-authors demonstrated that it is possible to achieve a very black coating of carbon nanotubes on a silicon substrate, establishing "the world's darkest material," which had an absolute absorptance inferred by reciprocity from reflectance measurements to have 0.9997 absorptance.[9] The work of Yang has captured the attention of the media and the race for something 'blacker than black' [Question for the editor to cite: "This is Spinal Tap," Rob Reiner, Embassy Pictures, 1982, VHS]. In the present context, something that is very black has very low reflectance (the lowest being less than 200 ppm) over a broad wavelength range. The absolute reflectance for nine different materials including nanotube samples are shown in Figure 2. A specific example of visible/near infrared absolute total hemispherical reflectance measurement method is described in Appendix 2.

Carbon nanotubes are not useful merely because one can demonstrate very low reflectance at a single wavelength. It is apparent there is nothing comparable to a vertically aligned nanotube array (VANTA) for being uniformly black over a very broad wavelength range, from the visible to far infrared (FIR). The electrical and thermal properties of carbon nanotube arrays are important when considering other types of black coatings and their advantages. However, a summary of electrical and thermal properties of carbon nanotubes is outside the scope of this review. Instead, the reader is directed to other reviews,[10] though we find that there is no definitive literature on this subject. Most investigations in the literature are highly specialized. There are hundreds of references on properties of nanotubes that would suggest that the basic knowledge of a multiwall nanotube array is known. The idea, however, of a 'textbook' value of, say, electrical conductivity, thermal conductivity, density, specific heat, or thermal diffusivity remains elusive. This is presumably because of the difficultly of measuring such properties of a deformable film (or layer) that is highly dependent on topology. For thermal detectors and blackbody emitters the high thermal diffusivity of vertically aligned carbon nanotubes is desirable.[11, 12, 13]

**A. Other Black Coatings**

This section provides a summary of the different types of black surfaces other than carbon nanotube-based black coatings. These surfaces are typically made black through anodization, chemical etching, spraying black paints or coatings onto them, or with optical interference coatings. Many of these coatings have been used successfully in high-performance scientific instruments; none, however, provide such low reflectance over as large a wavelength span as carbon nanotube surfaces. An excellent summary can be found in Pompea and Breault[14]. Reflectance spectra and aging properties of many coatings can be found in Dury et al.[15]

Black surfaces can be achieved through a combination of applying absorbing materials; creation of cones, holes, or cavities in the surface that are large relative to the wavelength of light; surface scattering; and optical interference. Including absorbing materials such as carbon soot in paints and dyes that are applied to surfaces is one way to create a black surface. There are numerous paints available commercially that incorporate carbon soot. Cones, holes, and cavities in the surface allow multiple reflections of the light within the surface and increase the probability of absorption. Commercial black anodization such as Martin Black[†] and Pioneer Optical Black[†] use the surface pitting of anodization and black dye to create a black surface. Robust and very black coatings without paints or dyes can be made by etching a nickel-phosphorus alloy to create a highly absorptive surface morphology (Figure 1).[16,18] Gold black and silver black achieve good absorption due to a micro-dendritic structure on the surface (Figure 1).[17] These coatings can be damaged by touch and performance degrades if the dendritic structures are damaged[18]. Multi-layer interference films, which are not dendritic and hence more robust, can also be used to achieve low reflectivity but over only relatively narrow wavelength ranges.

## II. THE PURPOSE AND NEED FOR BLACK COATINGS

Black surfaces are available in a wide variety of forms and are used for a diverse set of applications including solar thermal energy generation,[19] blackbodies,[20] thermal control, baffles and shrouds,[21, 22] and radiometers for measurement of power in optical wavelengths from the ultra-violet to THz. Solar thermal energy generation and radiometers use the high optical absorptivity of black coatings to convert optical energy to heat. In the case of solar thermal energy, the heat is then converted to electrical energy. In the case of radiometers, the heat is used to measure the optical power of the incident light. Baffles and shrouds use the high optical absorptivity of black surfaces for stray light control in optical systems. The black coating absorbs most of the light that impinges on it, greatly reducing the intensity of scattered light in, for example, an optical imaging system. Finally, since black surfaces that are highly absorptive correspondingly have a high emissivity, they are used to create high quality blackbody sources. Adibeykian et al. describe the IR emissivity of some black coatings widely used for this application: Nextel 811-21, Herberts 1534, Aeroglaze Z306, and Acktar Fractal Black.[19, 23]

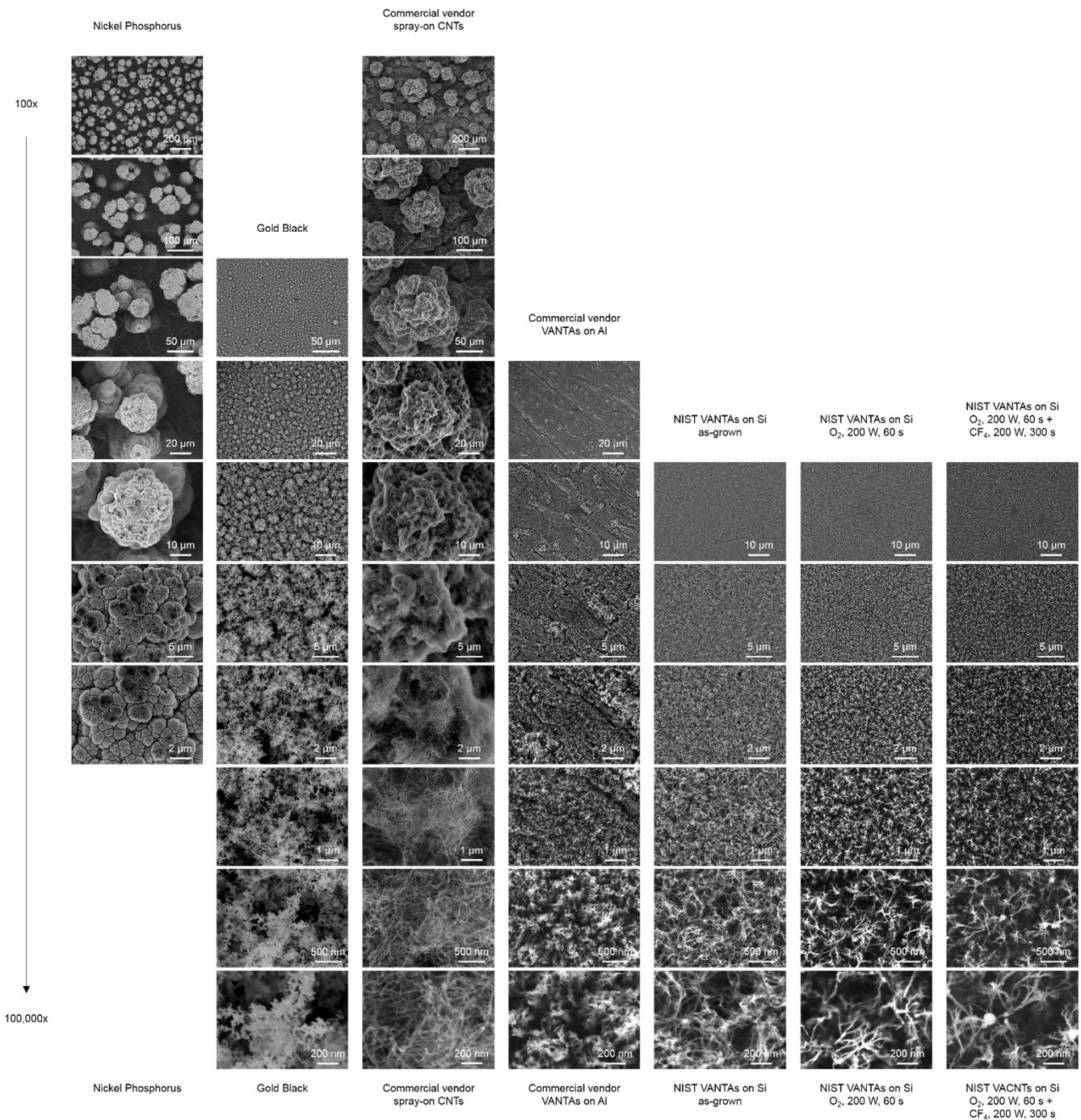

**Figure 1. FESEM images of black coatings measured in Figure 2.**

Black silicon, graphene, and tunable metasurfaces are newer approaches to creating black surfaces. As with the traditional black surfaces, these new varieties of absorbers do not exhibit the better than 0.1% absorptivity over the broad wavelength band and varied angles of incidence that carbon nanotubes offer. Black silicon suppresses reflection over a broad spectral range using a nanostructured surface etched onto the silicon surface to create silicon 'needles' that trap light through an effective medium process. Black silicon has been shown to reduce solar cell reflectivity from greater than 40 % to less than 2 % from 300 to 1000 nm.[24,25, 26]

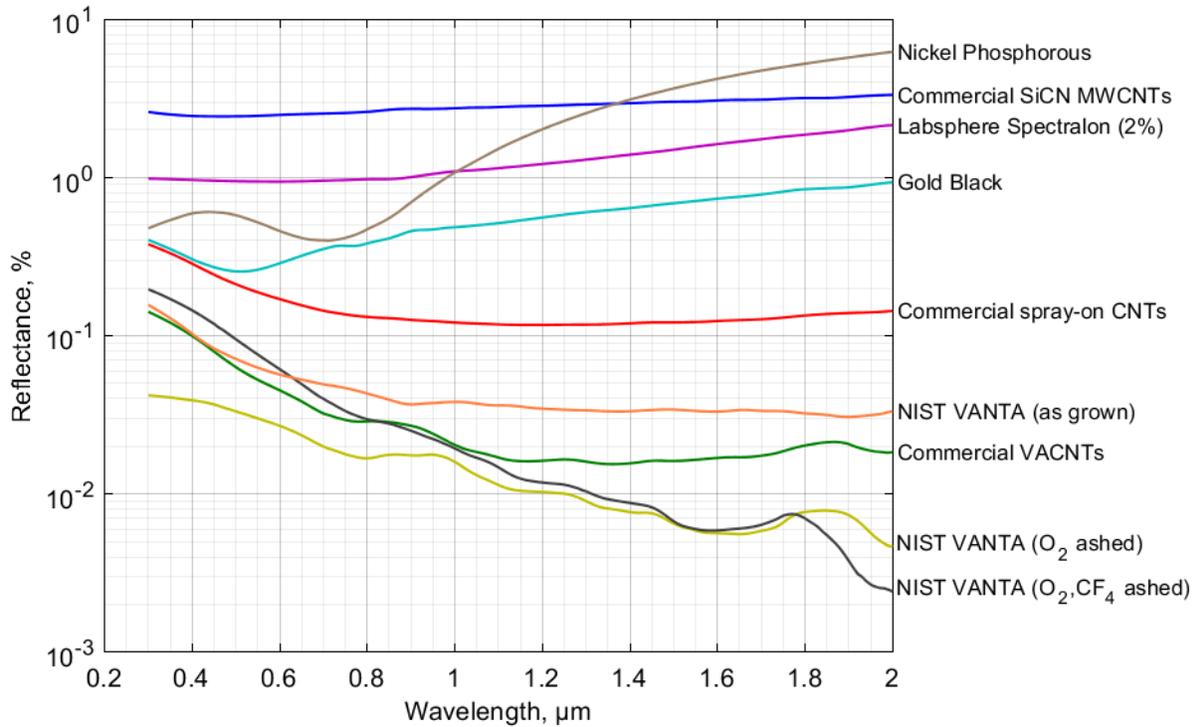

**Figure 2. Spectrophotometer measurements (see Appendix 2) showing the reflectance of six nanotube-based black coatings. A description of the measurement protocol is given in Appendix B.**

In the rapidly growing field of metasurfaces, artificial materials are created with nanoscale-engineered structures. Metamaterials are composed of subwavelength, periodic nanostructures that resonantly couple to the electromagnetic field.[27] Metasurfaces are single-layer or few-layer planar metamaterials that can be simpler to manufacture. The addition of sub-wavelength resonators to a surface interface modifies the boundary conditions by the resonant excitation of an effective current within the metasurface,[27] and allows the reflectance and transmission behavior to be tailored for specific applications, including creating tunable 'near-perfect' absorbers.[28, 29] The behavior depends on the wavelength of the incident light relative to the metasurface resonance. Metasurfaces have been combined with graphene, a one-atom-thick planar sheet of carbon atoms densely packed in a honeycomb crystal lattice to enhance absorption in the visible through the THz.[30, 31] Graphene's optical conductivity in the mid-infrared and THz frequency ranges is tunable by controlling the carrier density.[27]

### III.  SPACE QUALIFICATION

Carbon nanotube coatings are used in space applications such as stray light control,[32] bolometric measurements of Earth and Solar irradiance,[33] and blackbody sources.[20] Space missions that have used or plan to use carbon nanotube coatings include RAVAN[33] (GEOScan), ORCA,[32] and CIRiS.[34] Bolometers that use carbon nanotube absorbers for total and spectral solar irradiance measurements on CubeSats are under development. The harsh launch and on-orbit environments associated with space applications levy additional performance requirements on carbon nanotube coatings that are not typically considerations in the laboratory environment. These include survival in launch vibration environments, during shock, and in temperature extremes; adhesion on-orbit; and degradation of absorption in the presence of ultra-violet light, contamination, radiation, and atomic oxygen. The growth process and the process for the application of the coating as well as the surface preparation of the specific nanotube sample all impact the robustness of the coating to the space environment.

Theocharous et al.[35] performed outgassing, vibration, shock, and thermal cycle tests on VANTAs grown on aluminum substrates and found no significant mass loss and no significant reflectance change. Collins et al., had similar results for vibration testing and thermal cycling and additionally found no change in visible reflectance after exposure to radiation equivalent to 5 years in a 700 km sun-synchronous orbit with 5 mils aluminum equivalent shielding.[34] Lubkowski et al.[36] found no change in reflectivity of carbon nanotube forests after exposure to gamma irradiation equivalent to an estimated surface lifetime in geostationary orbit.

### A. Survival

Optically black carbon nanotube coatings must survive mechanical and thermal survival conditions. Adhesion of the nanotubes to the surface during vibration and shock is the primary mechanical concern. The adhesion of the nanotubes to the substrate depends on the details of the growth and application process. The growth catalyst and the surface roughness of the substrate contribute to the adhesion of VANTAs.[37,38] The use of a CNT paint with an epoxy binder, although not as black, may show better adhesion. If the coating does not adhere well the absorptivity may decrease, and the nanotubes may create particulate contamination. Particulate contamination can impact stray light, and if the particulate is conductive could create shorts. While still an area of research, to date no particulate contamination has been attributed to poor adhesion of carbon nanotube coatings (see Section VII, Health and Safety for a more detailed discussion).

Typical survival temperature ranges are -35 C to 70 C.[39] Carbon nanotube coatings that have been tested to date to similar or greater temperature ranges show no degradation after thermal cycling. [35,34]

### B. Ageing

Once on-orbit, exposure to the space environment may cause changes to the nanotube absorptivity or thermal conductivity. This includes changes in the optical absorption due to UV exposure, contamination, radiation exposure and atomic oxygen. Blue and Perkowitz measured a decrease in reflectance of six optical black coatings in the extreme infra-red after six years in space[40] but did not have a carbon nanotube sample.

Little direct work has been performed to determine long-term stability of the nanotube coating optical absorption due to UV exposure, contamination, and atomic oxygen. However, some conclusions can be drawn based on an understanding of the nanotubes. UV light and contamination go hand in hand. Certain types of contaminants have been shown to polymerize in UV and reduce the transmission on optical surfaces.[41] UV polymerization of contaminants on the surface of CNTs should be same mechanism as for optical components. Additionally UV light has been used to clean amorphous carbon off of single wall CNTs,[42] and photoinduced oxidation of CNTs measured via thermopower over a period of days[43] has been measured. The interaction of 248 nm light with a $\pi$ plasmon resonance in the CNTs when exposed to 248 nm UV light transfers energy to the CNTs and can result in oxidation of impurities on the surface. This process is much more efficient in air than in vacuum, so will not be as pronounced in a space mission, but the local atmosphere and contamination of the CNTs should be important considerations for space applications.

Space instruments in low Earth orbit may be exposed to atomic oxygen. While CNT coatings inside spacecraft or instruments are unlikely to be exposed to significant amounts of atomic oxygen, this could be an issue for baffles at instrument apertures. Jiao et al. showed that atomic oxygen exposure of carbon nanotube films result in mass loss and a changing degree of graphitization.[44] On earth (in the lab), oxygen plasma etching has been shown to reduce the reflectivity of vertically aligned carbon nanotube coatings.[45] Low energy (0.1 eV) oxygen plasma ashing in the lab is similar to actual atomic oxygen (4.5 eV) exposure in space.[46] Therefore, it is likely that long-term atomic oxygen exposure in space can modify the optical properties of carbon nanotube black coatings. It may be that surface treatments of carbon nanotubes, such as the addition of a $CF_4$ surface layer can reduce the impact of atomic oxygen in space applications.

The outgassing of carbon nanotube coatings is less likely to degrade the surface properties, but it is still important to measure in space as outgassing of the coatings may lead to contamination of optics or detectors near the coatings. CNTs are a low outgassing material, but it is important to remember that the outgassing properties depend on the details of the CNT coating. Table 1 shows the results of outgassing measurements on vertically aligned CNTs with different surface preparations. Black carbon nanotube coatings applied with an epoxy binder will be dominated by the outgassing of the binder. Theocharous et al.

also provides outgassing measurements and also includes RGA measurements that show no significant outgassing molecular species other than water, which is relative low.[35]

**Table I Outgassing measurements for samples of vertically alignment CNTs with different surface preparations. Outgassing is low, but varies for different types of CNT coatings.**

| Sample VANTA, 300-400 μm long | Total Mass Loss (TML) | Water Vapor Regained (WVR) | Collected Volatile Condensable Materials (CVCM) * |
|---|---|---|---|
| As grown | 0.006% | 0.007% | <0.004 % |
| Oxygen plasma ashed, 200 W, 60 s | 0.050% | 0.024% | < 0.006% |
| Oxygen plasma ashed, 200 W, 60 s + CF4, 30 W 60s | 0.055% | 0.019% | < 0.005% |

*The weight of the contaminants was under the limits of quantification (0.05 mg) for the test

**IV. HEALTH AND SAFETY**

Material considered nanoscale in more than one dimension raises caution with respect to human exposure. Carbon nanotube is one term for many possible morphologies. Like snowflakes, for example, no one has shown the world two identical carbon nanotubes, even among nominally similar growth conditions. Growing specific CNTs of known chirality and length and putting them exactly where they are needed remains a great challenge for the scientific community. A single, single-wall carbon nanotube could be nanometers in diameter and less than a micron in length. It could be straight or curved and helical. Meanwhile, a multiwall carbon nanotube could be 100 nm in diameter and many millimeters in length. Furthermore, many scientific studies have shown that it is extremely difficult to isolate a single nanotube. CNTs favor bundling, which is attributed to strong Van der Waals forces among tubes. The challenge of the bio researcher is several fold; knowing what she has, and repeatably incorporating the material in a manner that is not compromised by the use of surfactants or the extent of dispersion.

An important consideration that is qualitative and difficult to address are the routes of dispersal of nanotubes. For example, CNTs grown on a silicon substrate could be dislodged easily with a stylus, but readily survive shake tests and rinsing with water. Maynard and coauthors suggest that aerosol release of CNTs into air is low with handling.[47] A sprayed coating applied with a binder is a different matter. Such a coating is more like paint and if dislodged, might represent its own hazard depending on the binder.

We find in the literature, no studies that will explicitly point to nanotube toxicity generally. A well-known paper by Worlie-Knirsch and coauthors illustrated this problem in 2006.[48] A detailed review by Madani et al., sends a similar message; it is difficult to draw conclusions without very careful knowledge of the topology of the nanotubes, which is complicated by the fact that a single (monodispersed) nanotube is rare.[49] There is also the question of nano-sized metal catalysts and the likelihood that metal nanoparticles are responsible for biological effects as much as the nature of nanotubes. Pulskamp and coauthors discuss this.[50] Typically, nanotubes in bundles go beyond the nanoscale. Presently, CNTs (or CNT bundles) having an aspect ratio similar to that of asbestos are considered risky.[51] If single wall CNTs reach the lungs they are more toxic than carbon black or quartz.[52] In vitro investigation, however, indicate that SWCNTs are likely to be less toxic than carbon black and diesel exhaust particles.[53] Aschberger et al. conclude that the genotoxic potential of CNTs is currently inconclusive.[54]

## V. CARBON NANOTUBE GROWTH

### A. Catalyst

By the 1960s and possibly earlier, it was known that transition metals such as iron, cobalt, and nickel in the presence of carbon feedstock gases at elevated temperatures were key to the growth of 'flakes' of single crystal carbon.[55, 56] Early work by R. T. K. Baker had demonstrated the growth of filamentous carbon on transition metals[57, 58] with G. Tibbetts[59] recognizing in a review article the early discovery of carbon 'filaments' by Radushkevich and Lukyanovich,[60] now considered to be the first reported discovery of carbon nanotubes.[61] However, it is the discovery and report by S. Iijima[62] on 'helical microtubules of carbon' that spurred the rapid pace of research of carbon nanotubes over the past 20 years. Since then a variety of new catalysts[63] have been discovered along with various techniques to deposit and integrate carbon nanotubes into what is hoped may one day be a useful technology exploiting their extraordinary properties. An overview of the most widely used and conventional catalysts will be given along with a summary of more recently developed ones. For a more extensive and detailed review of catalytic chemical vapor deposition (CVD) of carbon nanotubes readers are referred to Dupuis[64] and Su et al.[65] for an overview of the growth mechanism of carbon nanotubes.

Conventional CVD growth of carbon nanotubes using hydrocarbon gases such as acetylene, ethylene, or methane, relies upon both a catalyst and a support catalyst layer to initiate growth of carbon nanotubes (non-catalytic growth is possible as was demonstrated by Iijima[62] and others[66] using carbon electrode arc-discharge). While researchers have developed a wide range of means by which to deposit thin film catalysts,[67-69] vacuum deposition remains the most popular due to repeatability and ease with which one can deposit the nanometer thick layers. Combined with the use of lift-off photolithography, the researcher can define selective regions of growth with sub-micrometer resolution. As found with early work growing filamentous carbon, thin layers of transition metals such as Fe, Ni, or Co give optimal growth results for carbon nanotubes. The as-deposited thin films form into nanoparticles upon heating (in a process known as particle coarsening or Ostwald ripening[70]) which act as individual carbon nanotube growth sites.

Iron catalyst layers, which may be the most commonly used, have been shown to support single-wall carbon nanotube (CNT) growth for any type of iron containing compound.[71] Other catalysts such as SiC,[63] gold nanoparticles,[72] solution deposited iron containing gels,[67] and core shell loaded ferritin[73, 74] have been demonstrated to be suitable as well. The catalyst layer thickness has a significant effect on the diameter,[75] number of walls,[76] growth rate, and areal density.[77, 78] Lieber et al. found that iron nanoparticle diameter precisely determines the nanotube diameter.[79] Similarly, Hata et al. showed careful control of nanoparticle size determined single wall CNT diameter using arc plasma deposition of nanoparticles.[80] It has also been found that how the catalyst layer is treated before growth has an effect as well.[81] Dai et al. showed that by loading controllable amounts of iron into a apoferritin core, one can control the resulting iron nanoparticle size after high temperature removal of the organic shell and the resulting nanotube diameter.[74]

As to why carbon nanotube growth is initiated on transition metals, research suggests that the growth mechanism for crystalline carbon on is due to the solubility of carbon at high temperatures into the metal aiding in vapor-liquid-solid (VLS) epitaxy.[82] TEM has been used to monitor the growth of single wall CNTs in real time on Ni particles. It was found that nanotubes initiated from Ni nanoparticles grew directly from carbon adsorbates rather than from an intermediate metal carbide such as $Ni_3C$.[83, 84] For iron catalysts, TEM analysis shows that the growth is mediated through an iron carbide.[85, 86]

To further enhance the catalytic activity of the metallic nanoparticles, a support catalyst is deposited in addition and before the catalyst layer. Again, as with iron, there is a preferred material for the support catalyst which in this case is aluminum oxide.[87] Other oxides/nitrides such as $SiO_2$,[88] MgO, TiN, and $ZrO_2$,[89] have been shown to work. In addition, the use of a support catalyst is highly sensitive to the material used, which crystalline face[90] is used if growing on sapphire, how it is deposited[91] and how it is treated post-deposition.[92]

Catalyst poisoning is another issue which limits the eventual height to which one can grow vertically aligned carbon nanotubes. It is believed that amorphous carbon growth leads to the eventual termination of the growth of the nanotubes on the metal catalyst site. However, this study questions whether amorphous carbon is the cause.[93] As a result, many researchers have developed means to enhance the catalytic activity of the surface upon which growth is initiated. One of the biggest discoveries regarding long length growth of carbon nanotubes may arguably be the discovery by Hata et al. of the use of water-assisted growth.[4] By simply bubbling an inert gas (He) through water, they were able to show 'super growth' of millimeter height SWNT in relatively short growth times (10 minutes). In contrast, Kawarada et al. were able to demonstrate centimeter height

SWNT growth using plasma enhanced CVD using growth times of 10s of hours.[94] Since the discovery of water-assisted growth, researchers have been trying to elucidate the mechanism by which water enhances the growth of CNTs. Maruyama et al. showed in a study that the addition of water inhibited Ostwald ripening[95] of Fe catalyst, an effect they believe to work in conjunction with the amorphous carbon etching. A number of researchers have shown similar results to that of water-assisted growth can be achieved by introducing oxygen.[96] Later Hata et al. further extended this understanding by demonstrating that the addition of any organic molecule (e.g. various alcohols, acetone, etc.) containing oxygen bonded to the carbon chain had similar effects as water on the enhanced catalytic activity.[97] And finally, in a suggestion that water injection provides a wider window for growth parameters, Noda et al. showed that water is not needed if the right ratio of gas is used.[98]

**B. Substrates**

Silicon is a popular choice because it has a high melting point of 1414 °C, is cheap and is a standard substrate for many cleanroom processes. Other substrates such as GaAs,[99] SiC,[100] LiTaO$_3$,[101] Al$_2$O$_3$,[90, 102, 103] SiO$_2$,[104] and MgO[105] are compatible as well but are more expensive and chosen mainly for a specific property of the substrate. III-V substrates have been considered for integration of CNTs with optoelectronics though this is not an area of high interest. Other lower temperature and lower cost glass substrates such as soda lime glass[106] (maximum mechanical service temperature 460 °C, softening point 760 °C) have been shown to work. In the end, the desire to eventually integrate carbon nanotubes and exploit their extraordinary properties has driven much of the research to integrate them with silicon electronics. This in turn has resulted in the development of carbon nanotube growth on compound substrates of thin film metals on silicon.[107]

Non-crystalline substrates such as metals have proven popular as well. Stainless steel,[108, 109] aluminum,[110, 111] Inconel,[112] and copper[113] are appealing since they are malleable substrates for non-planar objects such as cones or complicated light baffling. Stainless steel is currently being investigated for roll-to-roll production of CNTs.[114] Other flexible web materials such as polyimide film[115] (glass transition temperature of 350 °C) are compatible with very low temperature (200 °C) plasma enhanced CVD (PECVD) growth of VANTAs. In the end, any substrate is possible with CNTs that are applied by low temperature processes such as spin-coating,[116] spray-on CNTs,[117] or float-off VANTAs.[96]

**C. Tools and Temperatures**

The tools or systems to grow carbon nanotubes have become as varied as the types of catalyst used. We present a short summary detailing the most common method used for the growth of nanotubes, chemical vapor deposition. CVD is without doubt the most common method because of its low overhead in terms of equipment. Variants to the application of CVD such as the addition of a plasma, liquid injection (water[4] or catalyst[118]), catalyst pre-treatment,[81] gas composition,[119] heating methods,[111, 120] are current areas of ongoing research.

### 1. CVD

Chemical vapor deposition involves the flow of a reactive gaseous species over a substrate (catalytic or epitaxial) at elevated temperatures. How the gases are injected (pre-heating, flow dynamics, gas ratio, type of gas) and how to deliver non-gaseous species (solid carbon precursors, catalyst) are the main hurdles. Carbon nanotube growth benefits from not being dependent upon the wide range of carbon feedstock gases that can be used such as $CH_4$, $C_2H_2$, $C_2H_4$.[121] However, the relative partial pressures of the constituent gases has been shown to be critical to the catalyst lifetime.[98] System design is straight forward as a resistively heated quartz tube furnace is sufficient for almost any type of carbon nanotube growth. Cold-wall systems, which are typically a stainless steel chamber, allow for more flexibility. Complicated growth schemes such as the application of a DC bias grid and control of plasma placement (direct or remote) are easier to implement. Nonetheless, flow dynamics, location of the sample, pre-heating of the gases[122] have been found to play an important role in the growth mechanism.

### 2. PECVD

The use of plasmas to crack gaseous species into more reactive constituents during CVD growth is typically used to lower substrate growth temperatures.[123] In the case of CNT growth, the use of a plasma not only lowered the growth temperature to about 600 °C or even less, [124] but was also the most common method used in early demonstrations of vertically aligned growth.[106] The subsequent discovery of water-assisted growth[4] of vertically aligned single-walled nanotubes demonstrated that a plasma was not necessary for vertically aligned growth. Recent discoveries have even shown that something as simple as timed introduction of hydrogen into the CVD growth process of carbon nanotubes will transition from tangled to aligned,[122] or a sufficiently low partial pressure of $C_2H_2$ will achieve super-growth without water addition.[98] Meyyappan et al. provide excellent review articles regarding the history and application of PECVD.[125, 126]

Plasma enhanced CVD can take many forms, too numerous to detail here. In general, plasmas can be separated into two categories: a direct plasma (growth of CNTs occurs in the plasma generation region) and a remote plasma (radical formation is created up-stream from the substrate). Direct plasmas subject the carbon nanotubes to ion bombardment possibly resulting in etching or damage while remote plasmas minimize this. Direct or remote plasmas can be generated either using RF (MHz to GHz) or with high voltage DC. The different types of plasma type/generation may have various advantages or disadvantages but for the researcher it is often simply a matter of what is cheapest and most readily available. Remote plasmas can be generated inductively using a coil around a dielectric tube (quartz, alumina) through which the gas flows. Plasmas of this style are easy to integrate with tube furnace systems by placing the inductive coil upstream and outside the heated zone from the sample in the tube furnace. Direct plasmas require biasing a grid or electrode relative to the growth substrate. This configuration is easiest to implement in cold wall systems.

### D. Alignment

#### 1. Vertically aligned

Vertically aligned wafer-scale growth of carbon nanotubes was first demonstrated on sol-gel mesoporous silica.[104] Earlier demonstrations of aligned nanotubes were primarily limited to carbon arc-discharge synthesis and as a result were more practical for harvesting than for wafer-scale applications.[127, 128] Subsequent demonstrations of vertically aligned growth involved using specialized substrates such as anodized porous silicon[129] and plasma enhanced CVD.[106] Vertically aligned growth can now be achieved with: water injection, PECVD, correctly time hydrogen introduction[81]. For a more detailed description of the growth mechanism of VANTAs the reader is referred to Hart et al.[130] Since the first measurement of the extremely low reflectance of VANTAs,[9] additional measurements have been performed. An incomplete summary of published reflectances of VANTAs by various researchers is shown in Table I.

**Table II. Experimentally measured reflectance values for VANTAs.**

| VANTA reflectance (reported measurement) | Wavelength | Description (type, diameter, VANTA height, substrate; *not reported) | Year/Reference |
|---|---|---|---|
| 450 ppm (integrated total reflectance) | 633 nm | MWNT, 8-10 nm, 300 μm, * | 2008[9] |
| < 10,000 – 20,000 ppm (hemispherical reflectance (diffuse + specular)) | UV - NIR | SWNT, *, 300 – 500 μm, Si | 2009[11] |
| ≤ 1000 ppm | 600 nm | MWNT, *, 162 μm, $LiTaO_3$ | 2010[101] |
| 5800 ppm (directional-hemispherical reflectance) | 635 nm | MWNT, *, 166 μm, Si | 2010[131] |
| 5000 ppm (8°/h reflectance) | 600 – 700 nm | MWNT, 30 – 100 nm, 50 -100 μm, Si | 2010[132] |
| 500 ppm (reflectivity) | 635 nm | Interlinked MWNT, 40 nm, *, Si | 2012[133] |
| 180,000 ppm (reflectivity) | 635 nm | Aligned MWNT, > 100 nm, *, Si | 2012[133] |
| < 5000 ppm (total hemispherical (specular + diffuse)) | 600 nm | MWNT, 10 nm, 25 μm, Si | 2013[134] |

| 360 ppm (directional-hemispherical reflectance) | 700 nm | VANTA, *, 22 – 44 μm, Al | 2014[35] |

### 2. Matted

Matted or tangled-planar growth of carbon nanotubes was the initial growth orientation of CNTs until the discovery of methods to grow VANTAs. While VANTAs have displayed the lowest reflectances reported, matted CNTs are appealing because they are easily applied to substrates that are incompatible with high temperature CVD growth. Planar mats of spray-on SWNTs have been used as a black body absorber for a pyroelectric detector[135] and considered as a biomimetic moth-eye antireflection coating.[136] Interestingly, VANTAs that have been mechanically flattened into a 2D aligned mat show a reflectance of 10-15% (550-1150 nm).[137]

## VI. VANTA OPTICAL THEORY

Qualitatively, it is not surprising that VANTA are black since the morphology varies over a wide range of length scales and the nanotubes are typically quite sparse. To get a quantitative understanding, we review several theories aimed at modeling VANTA.

### A. Effective medium theory

Effective medium theories attempt to describe binary inhomogeneous media by averaging the electric fields of the two components to arrive at a homogeneous media with an effective dielectric function. The effective medium approximation (EMA) or Bruggeman's theory,[138] is typically used when the two components are comparable and neither one dominates the media. In the case of nanotubes, the EMA is appropriate for high densities of nanotubes in air or vacuum, as is the case for horizontally-aligned carbon nanotubes[139, 140] and spray-on nanotube coatings,[5, 141, 142] neither of which will be discussed here.

When one of the two components dominates, the Maxwell-Garnett approximation (MGA)[143] can be used, as is the case for vertically-aligned carbon nanotube arrays (VANTAs) (and some spray-on nanotube coatings[144]) where the nanotubes have a small fill factor in air or vacuum. Garcia-Vidal et al.[139] presented an effective medium theory based on the MGA to model VANTA. A schematic of the model is shown in Figure 3b, where individual carbon nanotubes are modeled as solid graphite rods. The ordinary ($\varepsilon_o$) and extraordinary ($\varepsilon_e$) dielectric functions of graphite (Fig. 3a) are discussed in Appendix C. The effective dielectric functions of the VANTA are [145, 146]

$$\varepsilon_{\text{eff}}^{\perp} = \frac{\sqrt{\varepsilon_e \varepsilon_o}(1+f) + \varepsilon_{\text{vac}}(1-f)}{\sqrt{\varepsilon_e \varepsilon_o}(1-f) + \varepsilon_{\text{vac}}(1+f)} \tag{1}$$

and

$$\varepsilon_{\text{eff}}^{\parallel} = f\varepsilon_o + (1-f)\varepsilon_{\text{vac}}, \tag{2}$$

where $\varepsilon_{vac} = 1$ is the vacuum dielectric function, $\varepsilon_{\text{eff}}^{\perp}$ ($\varepsilon_{\text{eff}}^{\parallel}$) is the component perpendicular (parallel) to the CNT direction and $f$ is the fill factor, which ranges from 0 to 1, defined as the volume fraction of graphite rods.

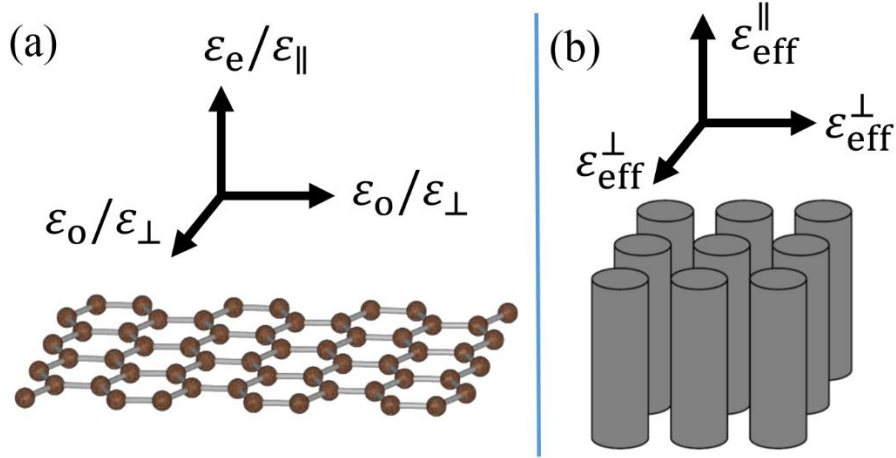

Figure 3. (a) Graphite model (see Section XII). Many references refer to the ordinary (extraordinary) dielectric function as the perpendicular (parallel) dielectric function, which is referring to alignment of the incoming light electric field with the surface normal (optical axis). (b) VANTA model of solid graphite rods. Here, perpendicular (parallel) is referring to the alignment of the incoming light electric field with the rod axis. Reprinted with permission from H. Bao, X. Ruan and T. S. Fisher, Opt. Exp. 18 (6), 6347-6359 (2010) [146]. Copyright 2010, Optical Society of America.

Imperfect alignment of the nanotubes can be accounted for by adding an alignment factor $x$, which can range from 0 to 1, where 1 is perfect vertical alignment and 0 is perfect horizontal alignment (all nanotubes lying flat on surface) as shown in Figure 4.[145] The VANTA dielectric functions are then[147]

$$\varepsilon_{\text{VANTA}}^{\perp} = x\varepsilon_{\text{eff}}^{\perp} + (1-x)\varepsilon_{\text{eff}}^{\parallel} \tag{3}$$

and

$$\varepsilon_{\text{VANTA}}^{\parallel} = x\varepsilon_{\text{eff}}^{\parallel} + (1-x)\varepsilon_{\text{eff}}^{\perp}. \tag{4}$$

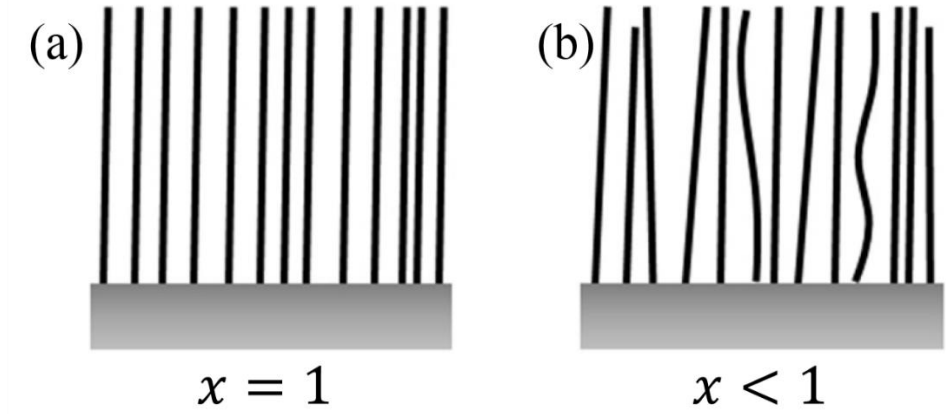

**Figure 4. Schematic showing (a) perfectly aligned and (b) imperfectly aligned VANTA. Reprinted with permission from R. Z. Zhang, X. Liu and Z. M. Zhang, Journal of Heat Transfer 137 (9), 091009-091009 (2015) [147]. Copyright 2015, American Society of Mechanical Engineers.**

If we define $\varepsilon_R$ and $\varepsilon_I$ as the real and imaginary parts, respectively, of the dielectric function, then the relation $N = (n + ik)^2 = \varepsilon_R + i\varepsilon_I$ can be used to calculate the VANTA index of refraction ($N$) real ($n$) and imaginary ($k$) parts. The effective medium theory has given us a dielectric function and index of refraction for VANTA, which can be used to model the optical properties.

## B. Thin film models

For simplicity, we first consider the reflectance of normal incidence light propagating from infinite medium of vacuum into an infinite medium of VANTA (Figure 5a). Reflectance off this single interface can be calculated using the Fresnel equations to get

$$R_{\text{Fresnel}} = \left|\frac{N_{\text{vac}} - N_{VANTA}}{N_{\text{vac}} + N_{VANTA}}\right|^2 \tag{5}$$

where $N_{vac} = 1$ is the vacuum index of refraction and $N_{VANTA}$ is the VANTA index of refraction only considering $\varepsilon_{\text{VANTA}}^{\perp}$ (Eq. 3). Figures 6 & 7 show theoretical curves of Fresnel reflectance for varying values of the fill factor $f$ and alignment $x$. The Fresnel reflectance is lowest at short wavelengths with sparse ($f \approx 0$) and well-aligned ($x \approx 1$) VANTA.

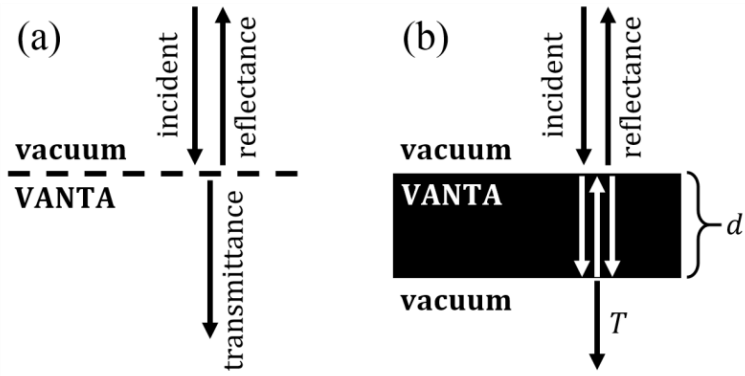

**Figure 5. (a) Schematic for Fresnel reflectance and transmittance from single vacuum/VANTA interface. (b) Schematic for reflectance and transmittance ($T$) from VANTA of thickness $d$.**

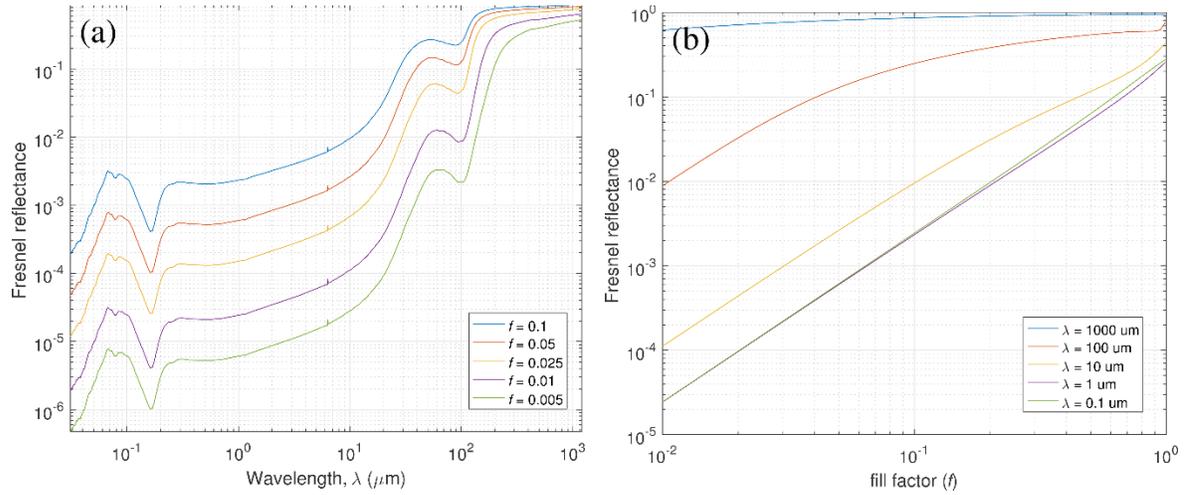

**Figure 6. Fresnel reflectance theory curves at a single vacuum/VANTA interface with alignment $x = 0.95$. (a) Plots vs. wavelength at various values of the fill factor $f$. (b) Plot vs. fill factor $f$ at various wavelengths.**

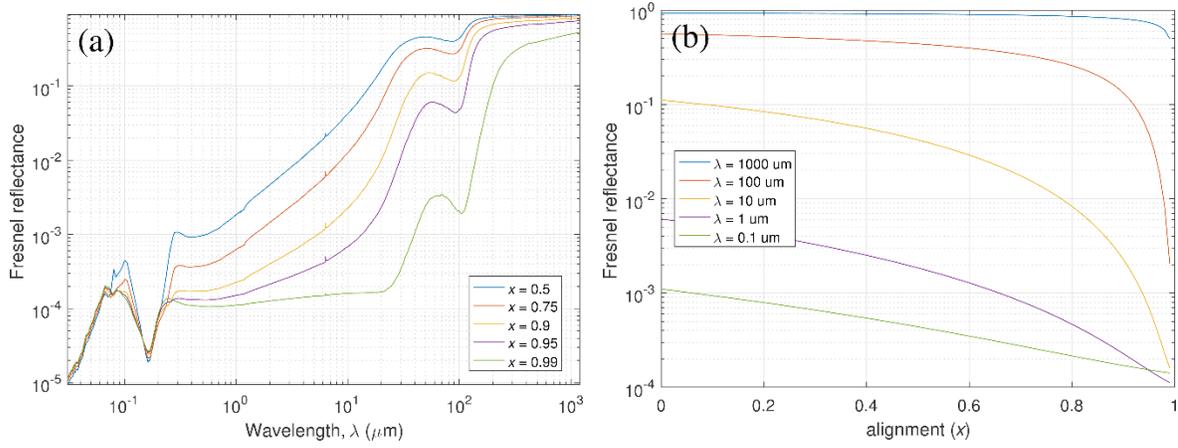

**Figure 7. Fresnel reflectance theory curves at a single vacuum/VANTA interface with fill factor $f = 0.025$. (a) Plots vs. wavelength at various values of alignment $x$. (b) Plot vs. alignment $x$ at various wavelengths.**

For a slightly more realistic model, we can use the transfer-matrix method to find the reflectance and transmittance for a finite thickness ($d$) of VANTA (Figure 5b). Figure 8 shows the reflectance and absorptance for different thicknesses of VANTA.

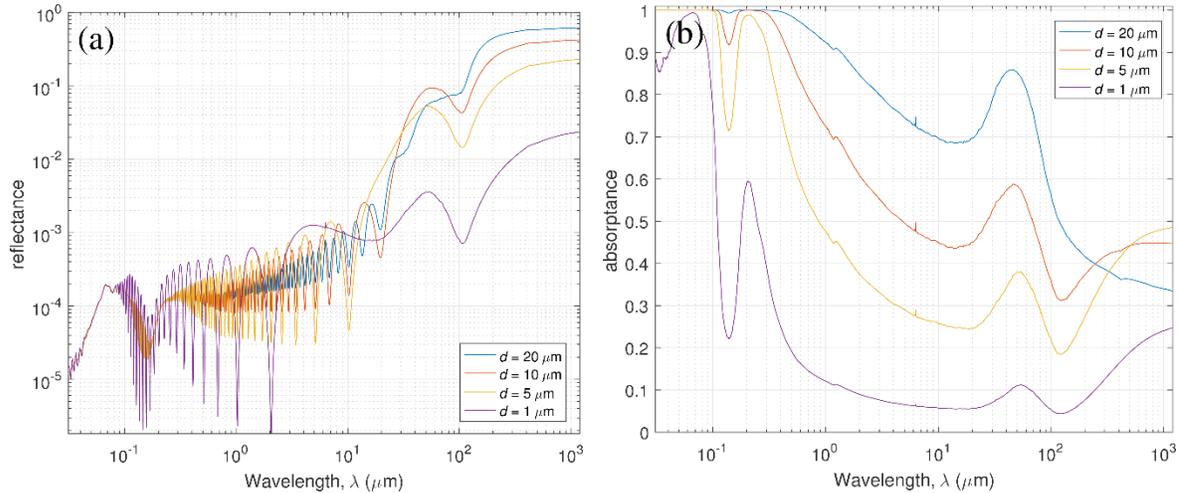

**Figure 8. Transfer-matrix theory curves for various thicknesses $d$ of VANTA films with a fill factor $f = 0.025$ and alignment $x = 0.95$. Plots are of (a) reflectance and (b) absorptance. Curves for thicknesses greater than $20\ \mu m$ are not shown because they look identical to the $20\ \mu m$ curve on the scale of the plots.**

Thin film models are quite simplified and have many limitations, one of which is that only specular reflection is calculated while diffuse reflection is absent because the interface is modeled as perfectly uniform. Several references[9, 148-150] expand the thin film model to account for surface roughness. The modeled film is also uniform, and therefore does not easily account for changes in VANTA density (fill factor) with height.

One alternative to the thin film model is a circular waveguide model proposed by Wood et al.[151]

## VII. MEASUREMENT OF LOW REFLECTANCE SAMPLES

In the past, national metrology institutes around the world have played an important role in the measurement of materials having low reflectance. This has been important for pure scientific interest as well as applied research and development of instrumentation. Measuring low reflectance can be accomplished in a direct or indirect way. In the direct case, this measurement involves detecting very small amounts of light reflected from the material as it is irradiated. This method is preferred for wavelengths from UV to mid infrared. The indirect approach is to measure emittance and then compute reflectance as the difference from unity (assuming the medium is opaque). This is preferred for longwave measurements beyond the mid infrared regime. For dark materials, it is crucial to minimize or correct typical sources of uncertainty including unwanted stray light and scattered light, thermal instability, over- or underfilling of the detector, detector nonlinearity and nonuniformity, drift of the measurement system, and others.

### A. Methods

In applied research, commercial systems are often used to initially determine the reflectance of black samples. The advantages of those instruments are short measurement times and ease of use but they often provide limited spectral resolution, moderate sensitivity and greatly varying levels of uncertainty between high and low reflectance samples. Most state-of-the-art measurement setups include some combination of the following components:

- Single wavelength (laser) or broad band light source (depending on the spectral range of interest, often a variable temperature blackbody, quartz tungsten halogen lamp, heated silicon carbide element, or mercury discharge lamp)
- Wavelength selector such as a Fourier transform interferometer or grating monochromator
- Vacuum or purge gas system
- Collector unit with detector (often an integrating sphere or hemi-ellipsoidal mirror with pyroelectric, HgCdTe, DLATGS based detector)
- Temperature controlled sample
- Reflectance reference standard (depending on the type of measurement, often a mirror, roughened metal, pressed powder, or thermoplastic polymer)

The light source is especially critical in the direct approach, and different types of measurements will be uniquely affected by its configuration. To achieve the highest signal-to-noise ratio and reduce measurement uncertainty it is always important to direct as much light as possible onto the sample under test.

There are many nuances involved in direct measurement of reflectance, which can depend on the wavelength range of interest as well as the reflectance of the sample. The character of imaging components (lenses or mirrors) must be known, and enclosures and baffles must be not transparent at the detector wavelength range. Any contributions from scattered light due to unwanted reflections from instrumentation components such as coatings or sharp edges must be measured and corrected for. Detector nonlinearity can also produce inaccurate results, particularly for a relative measurement if the absolute reflectance of reference standard is not comparable to the sample. For example, a highly reflective pressed-PTFE standard would not be an appropriate reference for measurement of a carbon nanotube coating unless the nonlinearity of the instrumentation is very well characterized and can be corrected. Often for very dark samples long averaging times are used to compensate for low signal, and therefore drift of the instrumentation over time becomes critical to the validity of the measurements. Multiple sample measurement cycles interspersed with reference measurements provide an advantage over one long measurement (additional labor notwithstanding), as the drift can be monitored and possibly corrected. Typically, but not always[150], carbon nanotube coatings are diffuse reflectors and therefore total hemispherical reflectance is the preferred metric. However, we expect the reflectance to become increasingly specular at longer wavelengths, which may adversely affect the measurement depending on the specific instrumentation used. Absorption and thermalization of the incoming light generates self-heating and the sample under test may have blackbody re-emission comparable to the light being reflected. See for example, descriptions by Betts[152] and later Chunnillal et al. [142] Uncorrected, this results in an offset; however, given that the two contributions are opposing in phase, they can be distinguished in the complex spectrum. The result is a lower, but more accurate reflectance. This effect can be significant even at relatively low temperatures (~ 100° above ambient).

These difficulties are often easier to overcome when using a relative rather than absolute measurement method. In a relative method, the sample is compared to a calibrated reference artifact. The relative reflectance of the sample with respect to that artifact is then scaled using the calibrated absolute reflectance of the artifact to yield the absolute reflectance of the sample. If the reflectance of the sample and standard artifact is similar, then some of the issues causing an increased uncertainty will cancel out between the sample and the reference measurements. If a well-calibrated reference is used, the benefit from this effect usually outweighs the added uncertainty of the absolute reflectance of the reference artifact. In the absolute case, all the contributing factors must be fully understood. The measurement apparatus must be carefully arranged, and in-depth empirical characterization is usually needed. This is typically preferred only if a high degree of accuracy is required. See Hanssen and Kaplan's[153] absolute system for diffuse reflectance measurements in the infrared for an example of the analysis required to properly perform a measurement using an absolute method.

It is worth clarifying that both absolute and relative measurement methods are used to determine the absolute reflectance of the sample. Despite its name, the relative method is not limited to measurement of relative reflectance because a standard artifact with known absolute reflectance is used as a reference. Commercial instruments for measuring directional hemispherical reflectance typically employ a relative method. The reference artifacts typically used with these instruments are available as commercial reflectance standards characterized over a range of wavelengths (often from UV to NIR), and are traceable to the PTFE primary reflectance standards[154, 155]. The role of the NMI has been to establish an SI traceable measurement of reflectance, which is typically realized by an absolute method used to directly determine the absolute reflectance of the primary reference materials.

### B. Means

Several authors[156] have written extensively on relevant spectroscopic methods, and it is not our aim to review these in detail; however, we will briefly mention several important works discussing reflectance and emittance measurement techniques. Betts, et al.[152] summarized a measurement method and results for five types of blacks in 1985. A measurement system for infrared hemispherical measurements was updated in 2012 by Chunnillal and Theocharous.[142] Hanssen describes a method and means for specular results.[157] Lindberg describes a method for deriving absolute diffuse reflectance results.[158] Laser sources don't provide broadband spectral information, but their high irradiance can give lower uncertainty reflectance measurements at discrete wavelengths. For example, see the laser reflectometer with integrating sphere and power meter operated at one wavelength by Yang et al.[9] Panagiotopoulos et al. also describe a laser reflectometer measurement system.[133] The National Metrology institute of Japan have several references for reflectance and emittance measurement systems [159,160] including an emittance measurement addressing the challenge of self-heating and re-emission from reflectance in the far infrared.[20] See also the integrating sphere instrumentation with FTIR (with uncertainty of 0.0058±0.0018 at 650nm) by Aschberger et al.,[54] and integrating sphere with FTIR and reference sample by Matsumoto et al.[134] Finally, the reader is directed to a book chapter on "Reflectance measurements of diffusing surfaces using conic mirror reflectometers" by Workman and Springsteen.[161]

As indicated in Section II, a detailed example of a direct UV to near infrared total hemispherical reflectance measurement using a relative method is described in Appendix 2. The results of this measurement for several black materials are shown in Figure 2.

**VIII. CONCLUDING REMARKS**

We have undertaken a review of carbon nanotube-based black coatings. The research continues to evolve along with the commercial availability and growth of applications including; thermal detector coatings, optical components, and coatings for baffles wherever stray light is unwanted. The most common method of production is chemical vapor deposition with variations in catalyst material, temperature, and substrate. Among the advantages of VANTA coatings is the exceptional broad, uniform and low reflectance. While the lowest reflection may help solve some technical problems, the method and means of measurement creates commensurate challenges in metrology and achieving uncertainties that are less than the inherent reflectance. It is possible to model the expected optical (and infrared) behavior of carbon nanotube-based coatings by an effective media approximation and other variations. We have summarized here the index of graphite from various sources in the literature and provide this as a useful tool in the supporting information. The nanoscale character of carbon nanotubes inherently attracts attention with respect to health and safety. We find that types and condition nanotubes, which are common to the blackest VANTA coatings do not pose a convincing risk, but precautions may be necessary for handling, particularly to prevent inhalation. Black coatings based on carbon soot have been pursued for more than one hundred years, but black coatings based on carbon nanotubes have been studied for less than fifteen years. We expect the existing work will provide a foundation for understanding the nature of black coatings generally and the understanding of carbon nanotubes.

**IX. ACKNOWLEDGEMENTS**

Thanks to Boris Wilthan for helpful discussions and guidance on measurement methods and instrumentation.

This manuscript is a contribution of the United States Government and is not subject to copyright.

[†] Use of tradenames is not an endorsement by NIST.

**X. APPENDIX A RECIPIES**

Here we provide recipes for spray-based and VANTA coatings. We recognize that this is an emerging technology and the best formulations may be proprietary (commercial) and therefore not found in the scientific literature.

### A. Simple Spray Coating

The basic formulation is as follows. The mass of nanotubes in solution may be varied depending on the length and purity of the tubes. Some tubes, poorly dispersed, will clog the airbrush. At the laboratory or experimental scale an artist's airbrush (Badger, Anthem Model 155[†]) is suitable. We have also used an industrial sprayer (Paasche, HVLP KRG-14[†]) for larger and more complicated geometries.[162]

1. 20 ml water
2. 0.5 ml potassium silicate
3. 50 mg nanotubes (more or less)
4. one 'flake' of DBS as surfactant if necessary

Mix the nanotubes and water (and surfactant) for about 10 minutes by means of a horn sonicator. Add the silicate and mix that for a couple minutes more (keeping the silicate cool). Place the substrate (aluminum in your case) on a hot plate anywhere between 60 and 90 C. A hot air gun (such as a hair dryer) can be useful for complicated shapes. Spray light passes. The initial coverage is not impressive, but in time it starts to pile up. Post bake at 300 C in an oven for two hours (or longer) to drive the water out of the silicate. Depending on the substrate temperature and nature of the tubes, one can achieve something more or less specular. This recipe is not optimized and we have not achieved lower than 5 % reflectance with this recipe, but the coating should be uniformly black over a broad wavelength range. Commercial (proprietary) formulations are much blacker as shown in Figure 2.

**B. Summary of PECVD Recipe**

Described below is a PECVD process used to grow VANTAs of multi-walled CNTs at NIST for black body absorbers. The PECVD growth chamber consists of a stainless-steel cold-wall vertical flow reactor with a carbon meander heater and a silicon wafer used as a susceptor (non-rotating). Heater temperature is monitored with a thermocouple mounted in contact with the heater element and not directly with the susceptor. As a result, substrate temperatures may be 100 – 200 °C less than the heater temperature. A remote plasma is generated above the substrate at a distance of approximately 30 – 35 cm (system is equipped with an adjustable height heater) using a waveguide coupled 2.45 GHz microwave source. Growth gases (Ar, $H_2$, $C_2H_4$) are injected via a quartz tube directly above the growth region which passes through the microwave plasma generation region (remote plasma configuration). The system is pressure controlled using a servo-actuated variable impedance valve and a roots style dry pump. Thin film catalyst is magnetron sputtered 10 nm of aluminum oxide (RF sputtered from an alumina target at 100 W in 0.40 Pa of Ar for 20 minutes) and 1 – 2 nm of iron (DC sputtered from a 0.25 mm thick Fe target using high strength rare-earth magnets at 50 W in 0.53 Pa of Ar for 1 minute). Thin film thickness is monitored in situ using a quartz crystal microbalance. Thermally oxidized silicon (100 – 200 nm) is an optimal substrate in terms of ease of growth. Other substrates, such as thin film metals on thermally oxidized silicon may require varying parameters such as growth temperature, thickness of support catalyst (in this case $Al_2O_3$), and possibly different support catalysts (refractory metals, metal nitrides, etc.).

1. Ramp to temperature: 800 °C at 50 °C/m in an Ar flow of 500 sccm at 2.666 kPa (20 Torr).
2. Growth: $H_2$ flow of 42.5 sccm (partial pressure of 0.227 kPa), $C_2H_4$ flow of 5 sccm (partial pressure of 0.027 kPa), Ar flow of 452.5 (partial pressure of 2.413 kPa), 900 W (2.45 GHz), total pressure of 2.666 kPa (20 Torr), growth rate ~ 250 μm/hour.
3. Cool down: Ar flow of 500 sccm at 2.666 kPa (20 Torr) until temperature is below 300 °C.

**C. VANTA Float-off transfer method**

Transfer of carbon nanotubes, whether vertically aligned or mat-like, is an appealing process as it eliminates subjecting the final device or substrate to high growth temperatures required for CNT formation. Two types of transfer methods exist for VANTAs. One utilizes an adhesive agent (such as glue or solder) on the final substrate to physically de-adhere and transfer the VANTAs from the growth substrate – a technique commonly used for thermal interface materials.[163] Another technique is truly a lift-off process whereby the VANTAs are released,[96, 163-166] picked up as a self-supporting substrate, and placed on the desired substrate which may or may not have an adhesive layer such as photoresist to reduce the effects of a thermal interface resistance. Since CVD grown VANTAs are hydrophobic,[167] the lift-off/float-off method lends itself well to the use of water-based etchants. Functionalized, plasma treated, or vacuum coated VANTAs which may be hydrophilic,[168, 169] are unsuitable for water-based lift-off methods as ingress of water into the forest will cause irreversible clumping of the forest. It has been shown that VANTAs may also be released without wet etchants by oxidation.[165, 166] We restrict our description of VANTA transfer utilizing hydrofluoric acid as a chemical etchant because of its ease and simplicity.

1. Place the VANTAs and substrate in a dilute (2-5% by vol.) solution of HF acid.
2. VANTAs will release from the substrate in a matter of seconds to minutes.
3. Transfer the VANTAs to multiple containers for HF removal.
4. Float the tubes onto the final substrate or lift them up using a flat surface (e.g. tweezers or a piece of flat plastic) and place them where desired.
5. Drying the transferred tubes requires minimal transfer of water as well as slow drying to prevent clumping and/or cracking. Ideally this should be done in a nitrogen dry box.

6. Oxygen plasma treat the nanotubes for a further reduction in reflectance.[45]

## XI. Appendix B – A description of a reflectance measurement method for Figure 2 as an example.

A commercial monochromator-based spectrophotometer with a 150 mm diameter integrating sphere accessory was used to obtain the directional-hemispherical reflectance spectra in Figure 2. The sphere has a rectangular entrance port and a 1 inch diameter sample mounting port, which are centered on the great plane of the sphere parallel to the optical bench. A photomultiplier tube and InGaAs detector are integrated into the bottom of the sphere wall and cover a wavelength range of 300–820 nm and 820–2000 nm respectively. The sphere wall is coated with a PTFE-based thermoplastic, which is highly reflective and diffuse in the UV, visible, and near-infrared[170].

Absolute reflectance values are determined by using a commercially available, calibrated reflectance standard made from a similar material to the sphere wall (although much darker). This artifact was chosen for its nearly-Lambertian scattering properties and low reflectance (roughly 5%), which reduce throughput difference and nonlinearity errors when comparing to samples which have similar properties. To obtain each reflectance spectrum, five distinct measurements are made in the following configurations:

1. sample mounted on the sphere and irradiated at 8° incidence,
2. reflectance standard mounted on the sphere and irradiated at 8° incidence,
3. sample port uncovered and beam passing through the sphere into an external beam dump
4. sample port uncovered and beam blocked before entering the sphere compartment
5. reflectance standard mounted on the sphere and the beam blocked before entering the sphere compartment

Note that all the following terms have implicit wavelength dependence. The signal produced by the photodetector in each configuration (minus dark signal) is directly proportional to the flux which irradiates the sphere, due to the highly reflective and nearly-Lambertian sphere coating[171]. In configurations 1 and 2, the photodetector measurements $V_1$ and $V_2$ are proportional to the sum of flux reflected from the sample or standard ($\phi_{refl,smp}$ or $\phi_{refl,std}$) and flux scattered from the entrance port of the sphere or overfilling the sample port ($\phi_{scat}$). $V_3$ is proportional to the sum of stray room light entering through the open sample port ($\phi_{room}$) and $\phi_{scat}$. $V_4$ is only proportional to $\phi_{room}$. $V_5$ is a measurement of the dark signal. Assuming system linearity, the signal component proportional to reflected flux from the sample and reference standard can be isolated with a linear combination of these measurements, shown by Equations B1 to B8.

$$V_{refl,smp} = V_1 - V_3 + V_4 - V_5, \tag{B1}$$

$$V_{refl,smp} = (V_{refl,smp} + V_{scat} + V_{dark} - V_{scat} - V_{dark} - V_{room} + V_{room} + V_{dark} - V_{dark}), \tag{B2}$$

$$V_{refl,smp} \propto \phi_{refl,smp} \tag{B3}$$

$$\phi_{refl,smp} = (\phi_{refl,smp} + \phi_{scat} - \phi_{scat} - \phi_{room} + \phi_{room}). \tag{B4}$$

For the standard's reflectance,

$$V_{refl,std} = V_2 - V_3 + V_4 - V_5, \tag{B5}$$

$$V_{refl,std} = (V_{refl,std} + V_{scat} + V_{dark} - V_{scat} - V_{dark} - V_{room} + V_{room} + V_{dark} - V_{dark}), \tag{B6}$$

$$V_{refl,std} \propto \phi_{refl,std}, \tag{B7}$$

$$\phi_{refl,std} = (\phi_{refl,std} + \phi_{scat} - \phi_{scat} - \phi_{room} + \phi_{room}). \tag{B8}$$

Thus, the ratio of $V_{refl,smp}$ and $V_{refl,std}$ is equal to the ratio of the reflected fluxes. If the incident flux is constant for the sample and standard measurements and instrument drift is minimal (which can be validated by interleaved sample and reference measurements), the value for the absolute reflectance of the sample can be found by multiplying the ratio of $V_{refl,smp}$ and $V_{refl,std}$ by the known reflectance of the standard:

$$\rho_{smp} = \frac{\phi_{refl,smp}/\phi_{incident}}{\phi_{refl,std}/\phi_{incident}} \rho_{std}, \tag{B9}$$

$$\rho_{smp} = \frac{V_{refl,smp}}{V_{refl,std}} \rho_{std}. \tag{B10}$$

The resulting reflectance spectrum can be especially noisy at wavelengths where the system responsivity is low. To improve readability in these regions, a LOESS smoothing algorithm is applied to the results. Linear local polynomials, the tricube weight function, and an $f$-value of 0.1 are used as described by Cleveland.[172] Expanded relative uncertainties ($k = 2$) in the measured spectra range from roughly 10% for the most reflective samples to about 60% for the least reflective: this is primarily due to deterioration of the signal-to-noise ratio of $V_1$ as the samples become more absorptive.

## XII. Appendix C – Graphite optical properties

For most models, the optical properties of graphite are needed for modeling the optical properties of carbon nanotubes. Since an in-depth review of graphite optical properties is beyond the scope of this work, we limit our attention to the most commonly used references for graphite in the carbon nanotube papers we have referenced. We have found that it is nontrivial to digitize the data and compare the different graphite references, so we attempt to do that here (see also supplementary information). References we compare are Draine 1984[173], Palik 1991[174], Kuzmenko 2008[175], and Zhang 2015[147]. Some references list the dielectric constants[173, 175] and some references list the index of refraction[147, 174]. For comparison, we have plotted all the references as index of refraction in Figures 9-11. For references that use the terminology perpendicular and parallel, we relabel as ordinary and extraordinary, respectively. Figure 11 shows the graphite index of refraction curves used for this work, which is mainly from Ref. [147], with Ref. [173] and Ref. [174] used to extend the range to shorter wavelengths. We use the relation $N = (n + ik)^2 = \varepsilon_R + i\varepsilon_I$ to derive the index of refraction from the dielectric function, and vice versa. The equations are:

$$\varepsilon_R = n^2 - k^2, \tag{C1}$$

$$\varepsilon_I = 2nk, \tag{C2}$$

$$n = \sqrt{(\varepsilon_R + \sqrt{\varepsilon_R^2 + \varepsilon_I^2})/2}, \tag{C3}$$

and

$$k = \varepsilon_I / 2n. \tag{C4}$$

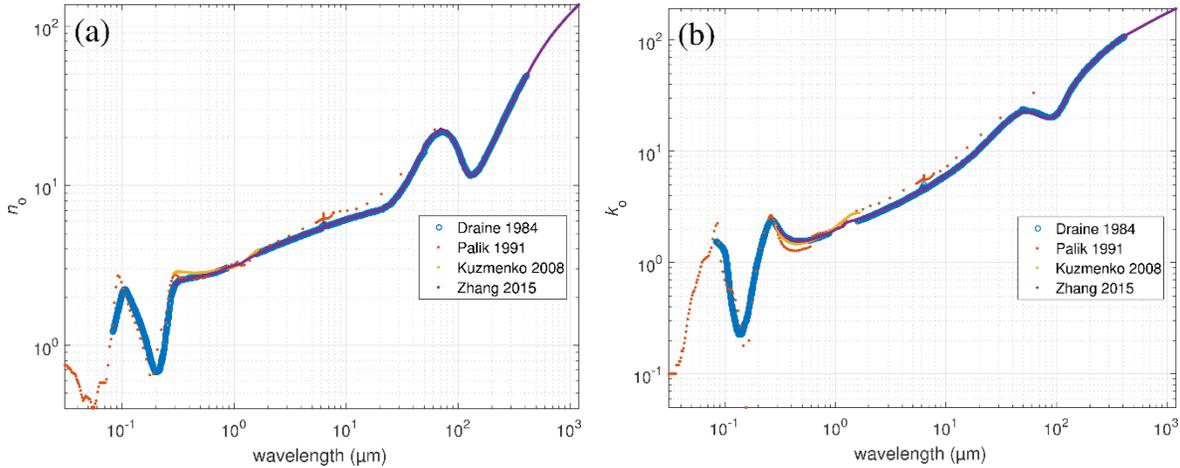

**Figure 9.** Plot of the ordinary index of refraction vs. wavelength for four different references. (a) Real part $n_o$. (b) Imaginary part $k_o$.

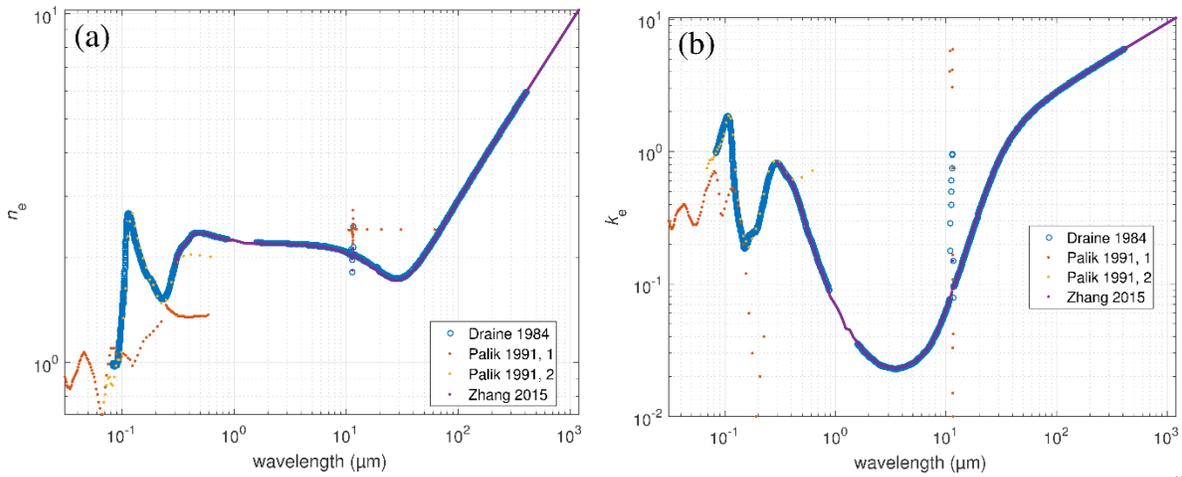

**Figure 10. Plot of the extraordinary index of refraction vs. wavelength for three different references. Palik[174] provides two separate lists, which we have labeled 1 and 2. (a) Real part $n_e$. (b) Imaginary part $k_e$.**

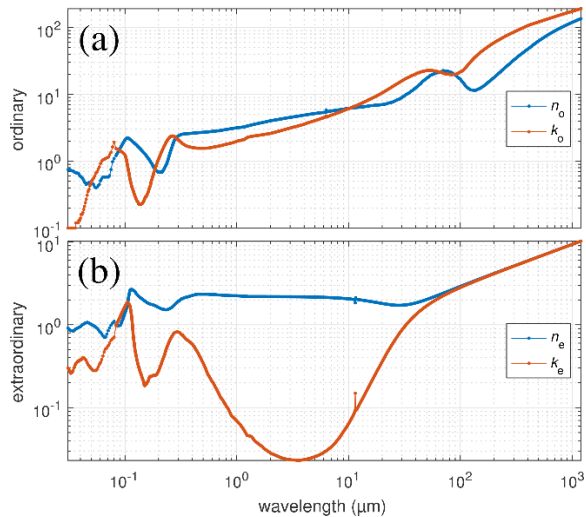

**Figure 11. Plots of the graphite index of refraction vs. wavelength used for this work. (a) Ordinary and (b) extraordinary.**

## REFERENCES

1. A. H. Pfund, J. Opt. Soc. Am. **23** (10), 375-378 (1933).
2. F. Kurlbaum, Annalen der Physik **301** (8), 746-760 (1898).
3. H. Servín, M. Peña, H. Sobral and M. González, Journal of Physics: Conference Series **792** (1), 012095 (2017).
4. K. Hata, D. N. Futaba, K. Mizuno, T. Namai, M. Yumura and S. Iijima, Science **306** (5700), 1362-1364 (2004).
5. J. H. Lehman, C. Engtrakul, T. Gennett and A. C. Dillon, Appl. Opt. **44** (4), 483-488 (2005).
6. F. F. Wang, Science (New York, N.Y.) **308** (5723), 838-841.

7. J. H. Lehman, R. Deshpande, P. Rice, B. To and A. C. Dillon, Infrared physics & technology **47** (3), 246-250 (2006).
8. E. Theocharous, R. Deshpande, A. Dillon and J. Lehman, Appl. Opt. **45** (6), 1093-1097 (2006).
9. Z.-P. Yang, L. Ci, J. A. Bur, S.-Y. Lin and P. M. Ajayan, Nano Letters **8** (2), 446-451 (2008).
10. R. Saito, G. Dresselhaus and M. S. Dresselhaus, *Physical Properties of Carbon Nanotubes*. (Imperial College Press, 1998).
11. K. Mizuno, J. Ishii, H. Kishida, Y. Hayamizu, S. Yasuda, D. N. Futaba, M. Yumura and K. Hata, Proceedings of the National Academy of Sciences **106** (15), 6044-6047 (2009).
12. J. H. Lehman, B. Lee and E. N. Grossman, Appl. Opt. **50** (21), 4099-4104 (2011).
13. M. Charan, L. L. Henry and W. Bingqing, Nanotechnology **16** (9), 1490 (2005).
14. R. P. B. Stephen M. Pompea, in *Handbook of optics*, edited by M. Bass and Optical Society of America. (McGraw-Hill, New York, 1995).
15. M. R. Dury, T. Theocharous, N. Harrison, N. Fox and M. Hilton, Optics Communications **270** (2), 262-272 (2007).
16. P. J. B. Richard J.C. Brouwn, Martin J.T. Milton, J. Mater. Chem. **12**, 2749-2754 (2002).
17. J. Lehman, E. Theocharous, G. Eppeldauer and C. Pannell, Measurement Science and Technology **14** (7), 916 (2003).
18. E. T. John Lehman, George Eppeldauer, and Chris Pannell, Meas. Sci. Tehcnol **14**, 916-922 (2003).
19. C. E. Kennedy, 2002.
20. K. Mizuno, J. Ishii, H. Kishida, Y. Hayamizu, S. Yasuda, D. N. Futaba, M. Yumura and K. Hata, Proc Natl Acad Sci U S A **106** (15), 6044-6047 (2009).
21. M. J. Persky, Review of Scientific Instruments **70** (5), 2193-2217 (1999).
22. T. Akutsu, Y. Saito, Y. Sakakibara, Y. Sato, Y. Niwa, N. Kimura, T. Suzuki, K. Yamamoto, C. Tokoku, S. Koike, D. Chen, S. Zeidler, K. Ikeyama and Y. Ariyama, Optical Materials Express **6** (5), 1613-1626 (2016).
23. A. Adibekyan, E. Kononogova, C. Monte and J. Hollandt, International Journal of Thermophysics **38** (6), 89 (2017).
24. V. E. Y. Howard M. Branz, Scott Ward, Kim M. Jones, Bobby To, Paul Stradins, Applied Physics Letters **94** (23), 231121 (2009).
25. S. Koynov, M. S. Brandt and M. Stutzmann, Applied Physics Letters **88** (20), 203107 (2006).
26. L. L. Ma, Y. C. Zhou, N. Jiang, X. Lu, J. Shao, W. Lu, J. Ge, X. M. Ding and X. Y. Hou, Applied Physics Letters **88** (17), 171907 (2006).
27. H. T. Chen, A. J. Taylor and N. Yu, Rep Prog Phys **79** (7), 076401 (2016).
28. K. Aydin, V. E. Ferry, R. M. Briggs and H. A. Atwater, Nature Communications **2**, 517 (2011).
29. Y. Avitzour, Y. A. Urzhumov and G. Shvets, Physical Review B **79** (4), 045131 (2009).
30. A. Andryieuski and A. V. Lavrinenko, Opt. Express **21** (7), 9144-9155 (2013).
31. S. Thongrattanasiri, F. H. L. Koppens and F. J. García de Abajo, Physical Review Letters **108** (4), 047401 (2012).
32. J. G. Hagopian, S. A. Getty, M. Quijada, J. Tveekrem, R. Shiri, P. Roman, J. Butler, G. Georgiev, J. Livas, C. Hunt, A. Maldonado, S. Talapatra, X. Zhang, S. J. Papadakis, A. H. Monica and D. Deglau, Proc. SPIE 7761, Carbon Nanotubes, Graphene, and Associated Devices III **77610F** (2010).
33. W. H. Swartz, L. P. Dyrud, S. R. Lorentz, D. L. Wu, W. J. Wiscombe, S. J. Papadakis, P. M. Huang, E. L. Reynolds, A. W. Smith and D. M. Deglau, 2015 IEEE International Geoscience and Remote Sensing Symposium (IGARSS), 5300-5303 (2015).
34. S. Collins, J. Fleming, B. Kelsic, D. Osterman and B. Staple, in *CalCon* (Logan, Utah, 2016).
35. E. Theocharous, C. J. Chunnilall, R. Mole, D. Gibbs, N. Fox, N. Shang, G. Howlett, B. Jensen, R. Taylor, J. R. Reveles, O. B. Harris and N. Ahmed, Opt Express **22** (6), 7290-7307 (2014).

36. G. Lubkowski, J. Kuhnhenn, M. Suhrke, U. Weinand, I. Endler, F. Meibner and S. Richter, IEEE Transactions on Nuclear Science **59** (4), 792-796 (2012).
37. J. G. Hagopian, S. A. Getty and M. A. Quijada, Patent No. US20130028829 A1 (2013).
38. N. Bahlawane, in *International Conference on Space Optics* (Biarritz, France, 2016).
39. NASA, 2013.
40. M. D. P. Blue, S., Appl. Opt. **31** (21) (1992).
41. J. Tveekrem, 1999.
42. K. E. Hurst, A. C. Dillon, S. Yang and J. H. Lehman, The Journal of Physical Chemistry C **112** (42), 16296-16300 (2008).
43. T. Savage, S. Bhattacharya, B. Sadanadan, J. Gaillard, T. M. Tritt, Y. P. Sun, Y. Wu, S. Nayak, R. Car, N. Marzari, P. M. Ajayan and A. M. Rao, Journal of Physics-Condensed Matter **15** (35), 5915-5921 (2003).
44. L. Jiao, Y. Gu, S. Wang, Z. Yang, H. Wang, Q. Li, M. li and Z. Zhang, Composites Part A: Applied Science and Manufacturing **71**, 116-125 (2015).
45. N. A. Tomlin, A. E. Curtin, M. White and J. H. Lehman, Carbon **74**, 329-332 (2014).
46. A. Banks, K. K. D. Groh, K. Rutledge, J. Difilippo, C. W. Reserve and T. International, (1996).
47. A. D. Maynard, P. A. Baron, M. Foley, A. A. Shvedova, E. R. Kisin and V. Castranova, J Toxicol Environ Health A **67** (1), 87-107 (2004).
48. J. M. Wörle-Knirsch, K. Pulskamp and H. F. Krug, Nano Letters **6** (6), 1261-1268 (2006).
49. S. Y. Madani, A. Mandel and A. M. Seifalian, Nano Reviews **4** (1), 21521 (2013).
50. K. Pulskamp, S. Diabate and H. F. Krug, Toxicol Lett **168** (1), 58-74 (2007).
51. K. Donaldson, C. A. Poland, F. A. Murphy, M. MacFarlane, T. Chernova and A. Schinwald, Advanced Drug Delivery Reviews **65** (15), 2078-2086 (2013).
52. C. W. Lam, J. T. James, R. McCluskey and R. L. Hunter, Toxicol Sci **77** (1), 126-134 (2004).
53. N. R. Jacobsen, G. Pojana, P. White, P. Moller, C. A. Cohn, K. S. Korsholm, U. Vogel, A. Marcomini, S. Loft and H. Wallin, Environ Mol Mutagen **49** (6), 476-487 (2008).
54. K. Aschberger, H. J. Johnston, V. Stone, R. J. Aitken, S. M. Hankin, S. A. Peters, C. L. Tran and F. M. Christensen, Crit Rev Toxicol **40** (9), 759-790 (2010).
55. P. L. Walker, J. F. Rakszawski and G. R. Imperial, J Phys Chem-Us **63** (2), 133-140 (1959).
56. D. Robertson, Carbon **8**, 365-374 (1970).
57. R. T. K. Baker, P. S. Harris, R. B. Thomas and R. J. Waite, J Catal **30** (1), 86-95 (1973).
58. R. T. K. Baker, Carbon **27** (3), 315-323 (1989).
59. G. G. Tibbetts, Journal of Crystal Growth **66** (3), 632-638 (1984).
60. L. Radushkevich and V. Lukyanovich, Zhurnal Fizicheskoi Khimii **26**, 88-95 (1952).
61. M. Monthioux and V. L. Kuznetsov, Carbon **44** (9), 1621-1623 (2006).
62. S. Iijima, Nature **354** (6348), 56-58 (1991).
63. M. H. Rummeli, A. Bachmatiuk, F. Borrnert, F. Schaffel, I. Ibrahim, K. Cendrowski, G. Simha-Martynkova, D. Placha, E. Borowiak-Palen, G. Cuniberti and B. Buchner, Nanoscale Res Lett **6** (1), 303 (2011).
64. A. C. Dupuis, Progress in Materials Science **50** (8), 929-961 (2005).
65. J. P. Tessonnier and D. S. Su, Chemsuschem **4** (7), 824-847 (2011).
66. T. W. Ebbesen and P. M. Ajayan, Nature **358** (6383), 220-222 (1992).
67. H. Kind, J. M. Bonard, L. Forro, K. Kern, K. Hernadi, L. O. Nilsson and L. Schlapbach, Langmuir **16** (17), 6877-6883 (2000).
68. H. C. Choi, S. Kundaria, D. W. Wang, A. Javey, Q. Wang, M. Rolandi and H. J. Dai, Nano Letters **3** (2), 157-161 (2003).
69. A. Gohier, K. H. Kim, E. D. Norman, L. Gorintin, P. Bondavalli and C. S. Cojocaru, Appl Surf Sci **258** (16), 6024-6028 (2012).

70. M. Bedewy, B. Viswanath, E. R. Meshot, D. N. Zakharov, E. A. Stach and A. J. Hart, Chem Mater **28** (11), 3804-3813 (2016).
71. S. Sakurai, H. Nishino, D. N. Futaba, S. Yasuda, T. Yamada, A. Maigne, Y. Matsuo, E. Nakamura, M. Yumura and K. Hata, J Am Chem Soc **134** (4), 2148-2153 (2012).
72. S. Bhaviripudi, E. Mile, S. A. Steiner, A. T. Zare, M. S. Dresselhaus, A. M. Belcher and J. Kong, Journal of the American Chemical Society **129**, 1516-1517 (2007).
73. J. M. Bonard, P. Chauvin and C. Klinke, Nano Letters **2** (6), 665-667 (2002).
74. Y. M. Li, W. Kim, Y. G. Zhang, M. Rolandi, D. W. Wang and H. J. Dai, Journal of Physical Chemistry B **105** (46), 11424-11431 (2001).
75. E. F. Kukovitsky, S. G. L'vov, N. A. Sainov, V. A. Shustov and L. A. Chernozatonskii, Chemical Physics Letters **355** (5-6), 497-503 (2002).
76. T. Yamada, T. Namai, K. Hata, D. N. Futaba, K. Mizuno, J. Fan, M. Yudasaka, M. Yumura and S. Iijima, Nat Nanotechnol **1** (2), 131-136 (2006).
77. M. Chhowalla, K. B. K. Teo, C. Ducati, N. L. Rupesinghe, G. A. J. Amaratunga, A. C. Ferrari, D. Roy, J. Robertson and W. I. Milne, Journal of Applied Physics **90** (10), 5308-5317 (2001).
78. Z. P. Huang, D. Z. Wang, J. G. Wen, M. Sennett, H. Gibson and Z. F. Ren, Applied Physics A **74** (3), 387-391 (2002).
79. C. L. Cheung, A. Kurtz, H. Park and C. M. Lieber, Journal of Physical Chemistry B **106** (10), 2429-2433 (2002).
80. G. Chen, Y. Seki, H. Kimura, S. Sakurai, M. Yumura, K. Hata and D. N. Futaba, Sci Rep **4**, 3804 (2014).
81. G. D. Nessim, A. J. Hart, J. S. Kim, D. Acquaviva, J. Oh, C. D. Morgan, M. Seita, J. S. Leib and C. V. Thompson, Nano Lett **8** (11), 3587-3593 (2008).
82. E. F. Kukovitsky, S. G. L'vov and N. A. Sainov, Chemical Physics Letters **317** (1-2), 65-70 (2000).
83. S. Helveg, C. Lopez-Cartes, J. Sehested, P. L. Hansen, B. S. Clausen, J. R. Rostrup-Nielsen, F. Abild-Pedersen and J. K. Norskov, Nature **427** (6973), 426-429 (2004).
84. M. Lin, J. P. Ying Tan, C. Boothroyd, K. P. Loh, E. S. Tok and Y. L. Foo, Nano Lett **6** (3), 449-452 (2006).
85. A. K. Schaper, H. Q. Hou, A. Greiner and F. Phillipp, J Catal **222** (1), 250-254 (2004).
86. H. Yoshida, S. Takeda, T. Uchiyama, H. Kohno and Y. Homma, Nano Lett **8** (7), 2082-2086 (2008).
87. S. Noda, K. Hasegawa, H. Sugime, K. Kakehi, Z. Y. Zhang, S. Maruyama and Y. Yamaguchi, Japanese Journal of Applied Physics Part 2-Letters & Express Letters **46** (17-19), L399-L401 (2007).
88. C. Mattevi, C. T. Wirth, S. Hofmann, R. Blume, M. Cantoro, C. Ducati, C. Cepek, A. Knop-Gericke, S. Milne, C. Castellarin-Cudia, S. Dolafi, A. Goldoni, R. Schloegl and J. Robertson, J Phys Chem C **112** (32), 12207-12213 (2008).
89. P. B. Amama, C. L. Pint, F. Mirri, M. Pasquali, R. H. Hauge and B. Maruyama, Carbon **50** (7), 2396-2406 (2012).
90. H. Hongo, M. Yudasaka, T. Ichihashi, F. Nihey and S. Iijima, Chemical Physics Letters **361** (3-4), 349-354 (2002).
91. P. B. Amama, C. L. Pint, S. M. Kim, L. McJilton, K. G. Eyink, E. A. Stach, R. H. Hauge and B. Maruyama, ACS Nano **4**, 895-904 (2010).
92. T. Tsuji, K. Hata, D. N. Futaba and S. Sakurai, J Am Chem Soc **138** (51), 16608-16611 (2016).
93. C. Schunemann, F. Schaffel, A. Bachmatiuk, U. Queitsch, M. Sparing, B. Rellinghaus, K. Lafdi, L. Schultz, B. Buchner and M. H. Rummeli, ACS Nano **5** (11), 8928-8934 (2011).
94. G. Zhong, T. Iwasaki, J. Robertson and H. Kawarada, J Phys Chem B **111** (8), 1907-1910 (2007).
95. P. B. Amama, C. L. Pint, L. McJilton, S. M. Kim, E. A. Stach, P. T. Murray, R. H. Hauge and B. Maruyama, Nano Lett **9** (1), 44-49 (2009).

96. G. Zhang, D. Mann, L. Zhang, A. Javey, Y. Li, E. Yenilmez, Q. Wang, J. P. McVittie, Y. Nishi, J. Gibbons and H. Dai, Proc Natl Acad Sci U S A **102** (45), 16141-16145 (2005).
97. D. N. Futaba, J. Goto, S. Yasuda, T. Yamada, M. Yumura and K. Hata, Adv Mater **21** (47), 4811-4815 (2009).
98. K. Hasegawa and S. Noda, ACS Nano **5** (2), 975-984 (2011).
99. R. Engel-Herbert, Y. Takagaki and T. Hesjedal, Mater Lett **61**, 4631-4634 (2007).
100. M. Kusunoki, M. Rokkaku and T. Suzuki, Applied Physics Letters **71** (18), 2620-2622 (1997).
101. J. Lehman, A. Sanders, L. Hanssen, B. Wilthan, J. Zeng and C. Jensen, Nano Lett **10** (9), 3261-3266 (2010).
102. H. Ohno, D. Takagi, K. Yamada, S. Chiashi, A. Tokura and Y. Homma, Jpn J Appl Phys **47** (4), 1956-1960 (2008).
103. S. Han, X. Liu and C. Zhou, J Am Chem Soc **127** (15), 5294-5295 (2005).
104. W. Z. Li, S. S. Xie, L. X. Qian, B. H. Chang, B. S. Zou, W. Y. Zhou, R. A. Zhao and G. Wang, Science **274**, 1701-1703 (1996).
105. B. Zheng, C. G. Lu, G. Gu, A. Makarovski, G. Finkelstein and J. Liu, Nano Letters **2** (8), 895-898 (2002).
106. Z. F. Ren, Z. P. Huang, J. W. Xu, J. H. Wang, P. Bush, M. P. Siegal and P. N. Provencio, Science **282** (5391), 1105-1107 (1998).
107. J. W. Ward, B. Q. Wei and P. M. Ajayan, Chemical Physics Letters **376** (5-6), 717-725 (2003).
108. L. M. Yuan, K. Saito, W. C. Hu and Z. Chen, Chemical Physics Letters **346** (1-2), 23-28 (2001).
109. C. Masarapu and B. Wei, Langmuir **23** (17), 9046-9049 (2007).
110. C. Emmenegger, P. Mauron, A. Zuttel, C. Nutzenadel, A. Schneuwly, R. Gallay and L. Schlapbach, Appl Surf Sci **162**, 452-456 (2000).
111. M. Ahmad, J. V. Anguita, V. Stolojan, T. Corless, J. S. Chen, J. D. Carey and S. R. P. Silva, Advanced Functional Materials **25** (28), 4419-4429 (2015).
112. S. Talapatra, S. Kar, S. K. Pal, R. Vajtai, L. Ci, P. Victor, M. M. Shaijumon, S. Kaur, O. Nalamasu and P. M. Ajayan, Nat Nanotechnol **1** (2), 112-116 (2006).
113. G. Li, S. Chakrabarti, M. Schulz and V. Shanov, J Mater Res **24** (9), 2813-2820 (2009).
114. R. Guzman de Villoria, A. J. Hart and B. L. Wardle, ACS Nano **5** (6), 4850-4857 (2011).
115. S. Hofmann, C. Ducati, B. Kleinsorge and J. Robertson, Applied Physics Letters **83** (22), 4661-4663 (2003).
116. J. W. Jo, J. W. Jung, J. U. Lee and W. H. Jo, ACS Nano **4** (9), 5382-5388 (2010).
117. R. Bhandavat, A. Feldman, C. Cromer, J. Lehman and G. Singh, ACS Appl Mater Interfaces **5** (7), 2354-2359 (2013).
118. H. W. Zhu, C. L. Xu, D. H. Wu, B. Q. Wei, R. Vajtai and P. M. Ajayan, Science **296** (5569), 884-886 (2002).
119. J. B. In, C. P. Grigoropoulos, A. A. Chernov and A. Noy, ACS Nano **5** (12), 9602-9610 (2011).
120. E. R. Meshot, D. L. Plata, S. Tawfick, Y. Zhang, E. A. Verploegen and A. J. Hart, ACS Nano **3** (9), 2477-2486 (2009).
121. H. Kimura, J. Goto, S. Yasuda, S. Sakurai, M. Yumura, D. N. Futaba and K. Hata, Sci Rep **3**, 3334 (2013).
122. A. J. Hart and A. H. Slocum, J Phys Chem B **110** (16), 8250-8257 (2006).
123. S. Hofmann, C. Ducati, J. Robertson and B. Kleinsorge, Applied Physics Letters **83** (1), 135-137 (2003).
124. S. Hofmann, C. Ducati, J. Robertson and B. Kleinsorge, Applied Physics Letters **83**, 135-137 (2003).
125. M. Meyyappan, J Phys D Appl Phys **42** (21) (2009).

126. M. Meyyappan, L. Delzeit, A. Cassell and D. Hash, Plasma Sources Sci T **12** (2), 205-216 (2003).
127. W. A. Deheer, W. S. Bacsa, A. Chatelain, T. Gerfin, R. Humphrey-Baker, L. Forro and D. Ugarte, Science **268** (5212), 845-847 (1995).
128. H. Jyun-Hwei, W.-K. Hsu and C.-Y. Mou, Advanced Materials **5** (9), 643-646 (1993).
129. S. S. Fan, M. G. Chapline, N. R. Franklin, T. W. Tombler, A. M. Cassell and H. J. Dai, Science **283** (5401), 512-514 (1999).
130. M. Bedewy, E. R. Meshot, H. C. Guo, E. A. Verploegen, W. Lu and A. J. Hart, J Phys Chem C **113** (48), 20576-20582 (2009).
131. X. J. Wang, L. P. Wang, O. S. Adewuyi, B. A. Cola and Z. M. Zhang, Applied Physics Letters **97** (16), 1-4 (2010).
132. J. J. Butler, G. T. Georgiev, J. L. Tveekrem, M. Quijada, S. Getty and J. G. Hagopian, Earth Observing Missions and Sensors: Development, Implementation, and Characterization **7862** (October), 78620D-78621 (2010).
133. N. T. Panagiotopoulos, E. K. Diamanti, L. E. Koutsokeras, M. Baikousi, E. Kordatos, T. E. Matikas, D. Gournis and P. Patsalas, ACS Nano **6** (12), 10475-10485 (2012).
134. T. Matsumoto, T. Koizumi, Y. Kawakami, K. Okamoto and M. Tomita, Opt Express **21** (25), 30964-30974 (2013).
135. J. H. Lehman, C. Engtrakul, T. Gennett and A. C. Dillon, Appl Opt **44** (4), 483-488 (2005).
136. F. De Nicola, P. Hines, M. De Crescenzi and N. Motta, Carbon **108**, 262-267 (2016).
137. T. Saleh, M. V. Moghaddam, M. S. M. Ali, M. Dahmardeh, C. A. Foell, A. Nojeh and K. Takahata, Applied Physics Letters **101** (6) (2012).
138. D. A. G. Bruggeman, Ann. Phys. **24**, 636 (1935).
139. F. J. García-Vidal, J. M. Pitarke and J. B. Pendry, Physical Review Letters **78** (22), 4289-4292 (1997).
140. W. Lü, J. Dong and Z.-Y. Li, Physical Review B **63** (3), 033401 (2000).
141. A. Ugawa, A. G. Rinzler and D. B. Tanner, Physical Review B **60** (16), R11305-R11308 (1999).
142. C. J. Chunnilall and E. Theocharous, Metrologia **49** (2), S73 (2012).
143. J. C. Maxwell Garnett, Philosophical Transactions of the Royal Society of London A **203**, 385 (1904).
144. K. Ramadurai, C. L. Cromer, L. A. Lewis, K. E. Hurst, A. C. Dillon, R. L. Mahajan and J. H. Lehman, Journal of Applied Physics **103** (1), 013103 (2008).
145. X. J. Wang, J. D. Flicker, B. J. Lee, W. J. Ready and Z. M. Zhang, Nanotech. **20**, 215704 (2009).
146. H. Bao, X. Ruan and T. S. Fisher, Opt. Exp. **18** (6), 6347-6359 (2010).
147. R. Z. Zhang, X. Liu and Z. M. Zhang, Journal of Heat Transfer **137** (9), 091009-091009 (2015).
148. G. Chen, Pennsylvania State University, 2003.
149. Z.-P. Yang, M.-L. Hsieh, J. A. Bur, L. Ci, L. M. Hanssen, B. Wilthan, P. M. Ajayan and S.-Y. Lin, Appl. Opt. **50** (13), 1850-1855 (2011).
150. X. J. Wang, L. P. Wang, O. S. Adewuyi, B. A. Cola and Z. M. Zhang, Applied Physics Letters **97** (16), 163116 (2010).
151. B. D. Wood, J. S. Dyer, V. A. Thurgood, N. A. Tomlin, J. H. Lehman and T.-C. Shen, Journal of Applied Physics **118** (1), 013106 (2015).
152. D. B. Betts, F. J. J. Clarke, L. J. Cox and J. A. Larkin, Journal of Physics E: Scientific Instruments **18** (8), 689 (1985).
153. L. M. Hanssen and S. Kaplan, Analytica Chimica Acta **380** (2), 289-302 (1999).
154. P. Y. Barnes, A. C. Parr and E. A. Early, Special Publication (NIST SP)-250-48 (1998).
155. V. R. Weidner and J. J. Hsia, J. Opt. Soc. Am. **71** (7), 856-861 (1981).

156. J. Workman and A. Springsteen, *Applied Spectroscopy: A Compact Reference for Practitioners*. (Elsevier Science, 1998).
157. L. Hanssen, Appl. Opt. **40** (19), 3196-3204 (2001).
158. J. D. Lindberg, Appl. Opt. **26** (14), 2900-2905 (1987).
159. J. Ishii and A. Ono, Aip Conf Proc **684** (1), 705-710 (2003).
160. J. Ishii and A. Ono, Measurement Science and Technology **12** (12), 2103-2112 (2001).
161. J. Jerry Workman and A. Springsteen, *Applied Spectroscopy: A Compact Reference for Practitioners*. (Academic Press, 1998).
162. K. Beloy, N. Hinkley, N. B. Phillips, J. A. Sherman, M. Schioppo, J. Lehman, A. Feldman, L. M. Hanssen, C. W. Oates and A. D. Ludlow, Physical Review Letters **113** (26), 260801 (2014).
163. A. Kumar, V. L. Pushparaj, S. Kar, O. Nalamasu, P. M. Ajayan and R. Baskaran, Applied Physics Letters **89** (16), 1-4 (2006).
164. S. Huang, L. Dai and A. W. H. Mau, Journal of Physical Chemistry B **103**, 4223-4227 (1999).
165. M. Wang, T. Li, Y. Yao, H. Lu, Q. Li, M. Chen and Q. Li, J Am Chem Soc **136** (52), 18156-18162 (2014).
166. L. J. Ci, S. M. Manikoth, X. S. Li, R. Vajtai and P. M. Ajayan, Advanced Materials **19** (20), 3300-+ (2007).
167. H. Li, X. Wang, Y. Song, Y. Liu, Q. Li, L. Jiang and D. Zhu, Angew Chem Int Ed Engl **40** (9), 1743-1746 (2001).
168. A. I. Aria and M. Gharib, Langmuir **27** (14), 9005-9011 (2011).
169. A. O. Lobo, S. C. Ramos, E. F. Antunes, F. R. Marciano, V. J. Trava-Airoldi and E. J. Corat, Mater Lett **70**, 89-93 (2012).
170. G. T. Georgiev and J. J. Butler, Appl. Opt. **46** (32), 7892-7899 (2007).
171. L. M. Hanssen, K. A. Snail and P. R. Griffiths, in *Handbook of Vibrational Spectroscopy* (John Wiley & Sons, Ltd, 2006).
172. W. S. Cleveland, Journal of the American Statistical Association **74** (368), 829-836 (1979).
173. B. T. Draine and H. M. Lee, Astrophysical Journal **285**, 89-108 (1984).
174. E. D. Palik, *Handbook of Optical Constants of Solids III*. (Academic Press, 1991).
175. A. B. Kuzmenko, E. Van Heumen, F. Carbone and D. Van Der Marel, Physical Review Letters **100** (11), 2-5 (2008).